\documentclass[pra,reprint,amsmath,amssymb,showpacs,twocolumn]{revtex4-1}

\usepackage[dvipdfmx]{graphicx}
\usepackage{bm}
\usepackage{mathrsfs}
\usepackage{color}

\usepackage{ulem}
\newcommand{\revise}[2]{\textcolor{blue}{\sout{#1}} \textcolor{magenta}{#2}}
\renewcommand{\revise}[2]{\textcolor{black}{#2}}
\newcommand{\comment}[1]{\textcolor{green}{\bf 【{#1}】}}
\renewcommand{\comment}[1]{}

\begin{document}
\title{Axion instability near topological
quantum phase transition point}
\title{Axion instability and non-linear electromagnetic effect}
\author{Tatsushi Imaeda}
\affiliation{Department of Applied Physics, Nagoya University, Nagoya 464-8603, Japan}
\author{Yuki Kawaguchi}
\affiliation{Department of Applied Physics, Nagoya University, Nagoya 464-8603, Japan}
\author{Yukio Tanaka}
\affiliation{Department of Applied Physics, Nagoya University, Nagoya 464-8603, Japan}
\author{Masatoshi Sato}
\affiliation{Yukawa Institute for Theoretical Physics, Kyoto University, Kyoto 606-8502, Japan}


\date{\today}

\begin{abstract}
We investigate the instability due to dynamical axion field near
the topological phase transition of insulators.
We first point out that
the amplitude of
dynamical axion field is bounded for magnetic
insulators in general, which suppresses the axion instability.
Near the topological phase transition, however, the axion
field may have a large fluctuation, which decreases the critical
electric field for the instability and increases the axion induced
magnetic flux density.
Using two different model Hamiltonians, we report the electromagnetic
response of the axion field in details.

\end{abstract}
\maketitle

\cleardoublepage
\pagenumbering{roman}
\cleardoublepage
\newpage
\pagenumbering{arabic}


\section{Introduction}
Instability is a key to explore new phenomena in physics.
After instability, quantum states are rearranged and a new state of matters appears.
A well-known example is the Cooper instability, in which paired electrons condensate and the system hosts superconductivity\cite{Cooper}. 

In the context of topological phases of matters, roles of the topological $\theta$-term
in condensed matter physics have been discussed recently\cite{Essin,Qi09,Karch,Franz10,Bermudez10,Vazifeh10,Lan,Nomura11,Sekine14,Wang15,Morimoto15,
Vazifeh13,Kim14,Burkov15,Tse10,Maciejko10,Lan11,Mal'shukov13,Wu1124,Zhong15,Li10,Ooguri12,
Lee15,Wang16,Sekine16,Wang13,You16,Akamatsu,Shiozaki,Goswami}.
Such a topological term provides nontrivial phenomena which have not been observed in ordinary materials.
Modifying the Maxwell equation, the $\theta$-term in topological insulators gives
the quantized Kerr effect\cite{Tse10,Maciejko10,Lan11,Mal'shukov13,Wu1124,Zhong15} 
and the quantized topological magnetoelectric effects\cite{Essin,Qi09,Karch,Franz10,Bermudez10,Vazifeh10,Lan,Nomura11,Sekine14,Wang15,Morimoto15}.
Furthermore, magnetic monopoles can be realized as mirror images of electrons due to the Witten effect\cite{Qi09}.

A new instability arises due to the $\theta$-term when it fluctuates.
When $\theta$ is dynamical, which we call axion, it couples to electromagnetic fields,
changing the behaviors of electromagnetic propagating modes drastically\cite{Li10}. 
Using an analogy between particle physics and condensed-matter one, 
the axion fluctuation is shown to induce an instability when an applied electric field exceeds a critical value. 

A subtlety of the above analysis is that it uses $\theta$ itself as a dynamical variable. 
Being different from particle physics, no direct axion field exists in condensed-matter systems. 
As was shown in Ref.\cite{Li10}, an antiferromagnetic fluctuation can induce fluctuations of $\theta$,
but the correspondence between them is not exact.
In actual condensed-matter physics, the antiferromagnetic field, not the axion field, is primary.
Therefore, to understand the instability microscopically, the analysis should be re-examined in terms of the primary antiferromagnetic field. 

In this paper, we examine the axion instability in a microscopic point of view.
We introduce an antiferromagnetic field, instead of {an} axion field, and analyze the instability caused by the antiferromagnetic field.
In contrast to the naive expectation,
it is found that fluctuations of the antiferromagnetic field are insufficient to induce a detectable axion instability.
We reveal that $\theta$ is bounded above as a function of the antiferromagnetic field, 
and thus {its effect on electromagnetic fields} is strongly suppressed.
In order to enhance the axion {field}, fluctuations other than the antiferromagnetic field is necessary.
Analysis of quantum anomaly implies that the necessary fluctuation is related to topological quantum phase transition.
Only when fluctuations of the antiferromagnetic order and the topological quantum phase transition coexists,
\revise{the axion instability occurs.}{the induced axion field causes significant effects.}

\revise{This paper is organized as follows:      
In Sec.~\ref{sec:review}, we review the axion instability discussed by Ooguri and Oshikawa\cite{Ooguri12},
After that, we emphasize both magnetic order and  topological phase parameter are important for the axion instability.
In Sec.~\ref{sec:lowenergy} we consider the low energy effective theory of the axion electrodynamics.
In Sec.~\ref{sec:calculation} we consider a certain setup and calculate  static solutions and the critical electric field of axion instability
In Sec.~\ref{sec:model} we analyze the axion instability based on the model potential in two cases.
In Sec.~\ref{sec:summary} We close with a discussion and summary.}
{
This paper is organized as follows.
In Sec.~\ref{sec:review}, we start from reviewing the axion instability discussed by Ooguri and Oshikawa\cite{Ooguri12},
where the axion field is described in terms of the antiferromagnetic order.
We then qualitatively discuss how the non-magnetic fluctuations,
which are related to the topological order, affect the axion instability and resulting electromagnetic fields.
The detailed calculations including the non-magnetic fluctuations are given in Secs.~\ref{sec:lowenergy}-\ref{sec:appliedB}.
Section \ref{sec:lowenergy} derives the low energy effective theory of the axion electrodynamics
from a microscopic model of a topological insulator.
Section~\ref{sec:calculation} gives general forms of static solutions for the induced fields in a certain setup
under an applied electric field.
The critical electric field for the instability is also given in this section.
Section~\ref{sec:model} analyzes two model cases with and without the non-magnetic fluctuations.
Section~\ref{sec:appliedB} discusses the case when a magnetic field, instead of an electric field, is applied.
Section~\ref{sec:summary} summarizes the paper.
}


\section{Axion Instability due to Magnetic and Non-magnetic Fluctuations}
\label{sec:review}
\subsection{Review of Ooguri-Oshikawa's theory}
\label{sec:OO}
We first briefly review the axion instability discussed by Ooguri and Oshikawa\cite{Ooguri12}.
They start from the axionic electrodynamics in an insulator.
The effective Lagrangian density {is given by} 
${\mathcal L}={\mathcal L}_{\rm em}+{\mathcal L}_{\rm a}+{\mathcal L}_{\rm \theta}$, where
\begin{align}
&{\mathcal L}_{\rm em}=\frac{1}{8\pi}\left(\epsilon {\bm E}^2-\frac{1}{\mu}{\bm B}^2\right),\\
&{\mathcal L}_{\rm a}=g^2J\left[
(\partial_t\delta\theta)^2-\nu_{i}^2(\partial_i\delta\theta)^2-m^2(\delta\theta)^2\right],
\label{eq:axion}\\
&{\mathcal L}_{\rm \theta}=\frac{\alpha}{4\pi^2}(\theta_0+\delta\theta){\bm E}\cdot{\bm B},
\end{align}
are the Lagrangian densities of the electromagnetic fields $\bm E$ and $\bm B$, the axion field  
$\theta= \theta_0+\delta\theta$, and \revise{their}{the} interaction between them, respectively,
with $\epsilon$ and $\mu$ being \revise{}{the} dielectric constant and \revise{}{the} magnetic permeability, \revise{}{respectively,}
and $\alpha=e^2/\hbar c$ the fine-structure constant.
Here, $\mathcal{L}_{\rm a}$ originates from the Lagrangian density of the spin-wave mode in the insulator:
In a linear approximation, the fluctuation of the axion field $\delta\theta$ is related to that of the antiferromagnetic order $\delta\phi_5$ as
\begin{eqnarray}
\delta\theta=\delta \phi_5/g,
\label{eq:theta-phi5}
\end{eqnarray}
with $g$ being \revise{}{a} constant, and $J$, $\nu_i$ and $m$ are the stiffness, velocity and mass of the spin-wave mode, respectively.
By solving the equation of motion for $\delta\theta$ under an external  uniform electric field $E_0$,
they found that the system is unstable when the electric field exceeds the critical value given by
\footnote{
Note that the critical electric field $E_{\rm crit}$ in Ref.\cite{Ooguri12} is given by that inside the axion insulator,
which satisfies $\epsilon E_{\rm crit} =\epsilon_0 E_0^{\rm crit}$}
\begin{eqnarray}
E_0^{\rm crit}=\frac{m\epsilon}{\alpha\epsilon_0}\sqrt{\frac{(2\pi)^3g^2 J}{\mu}},
\label{eq:Ecritical_OO}
\end{eqnarray}
with $\epsilon_0$ being \revise{}{the} dielectric constant of the external material.
They further showed by solving the modified Maxwell equation derived from the effective Lagrangian $\mathcal{L}$
that the instability leads to \revise{complete screening of the electric field}{screening of the excess electric field above $E_0^{\rm crit}$}
and induction of a magnetic flux density inside the insulator.
\revise{}{(See Appendix~\ref{sec:ooguri-oshikawa} for the details.)}
In such a situation, however, the axion field becomes much larger than unity and Eq.~\eqref{eq:theta-phi5} no longer holds.
We therefore need to re-examine the relation between the axion field and the antiferromagnetic order beyond the linear approximation.

\subsection{Effect of the non-magnetic \revise{fluctuation}{fluctuations}}
\label{sec:nonmag_fluctuation}
Although the fluctuations of the antiferromagnetic order \revise{is}{are} necessary for the axion instability as discussed in Refs.\cite{Li10,Ooguri12}, 
we note that the antiferromagnetic order is not the only order relevant to the axion field. 
In general, \revise{the}{an} axion field $\theta$ can be regarded as a phase of a complex field $\phi=\rho e^{i\theta}$ $(\rho=|\phi|)$. 
Denoting the real (imaginary) part of $\phi$ as $\phi_4$ ($\phi_5$), 
i.e., $\phi=\phi_4+i\phi_5$, we find that $\phi_4$ and $\phi_5$ transform as
\begin{eqnarray}
\phi_4\rightarrow \phi_4,
\quad
\phi_5\rightarrow -\phi_5, 
\end{eqnarray}
under time-reversal and inversion operations, 
because $\theta$ transforms as $\theta\rightarrow -\theta$ under each of these transformations.
[Note that the symmetry of $\theta$ is determined so \revise{as}{that} the $\theta$-term ($\mathcal L_\theta\propto\theta {\bm E}\cdot{\bm B}$)
\revise{to be}{is} invariant under these transformations.]
Since $\phi_4$ and $\phi_5$ have different symmetry properties, they generally correspond to different orders in an insulator.
Indeed, while $\phi_5$ represents an antiferromagnetic order that breaks both the time-reversal and inversion symmetries, as we expected,
$\phi_4$ corresponds to a non-magnetic order that \revise{preserving}{preserves} both these symmetries.
This is sharply contrast to the axion field in particle physics.
In particle physics, the system has a U(1) axial symmetry that relates $\phi_4$ and $\phi_5$ to each other (so-called Peccei-Quinn symmetry\cite{Peccei}).
Therefore, even when $\phi_5$ takes a non-zero expectation value,
time-reversal and inversion symmetries can be retained by combining them with the U(1) symmetry.
Thus, these two fields represent essentially the same order in the particle physics.
On the other hand, no such a U(1) symmetry exists in magnetic insulators.

\begin{figure}[h]
\begin{center}
\includegraphics[width=5cm]{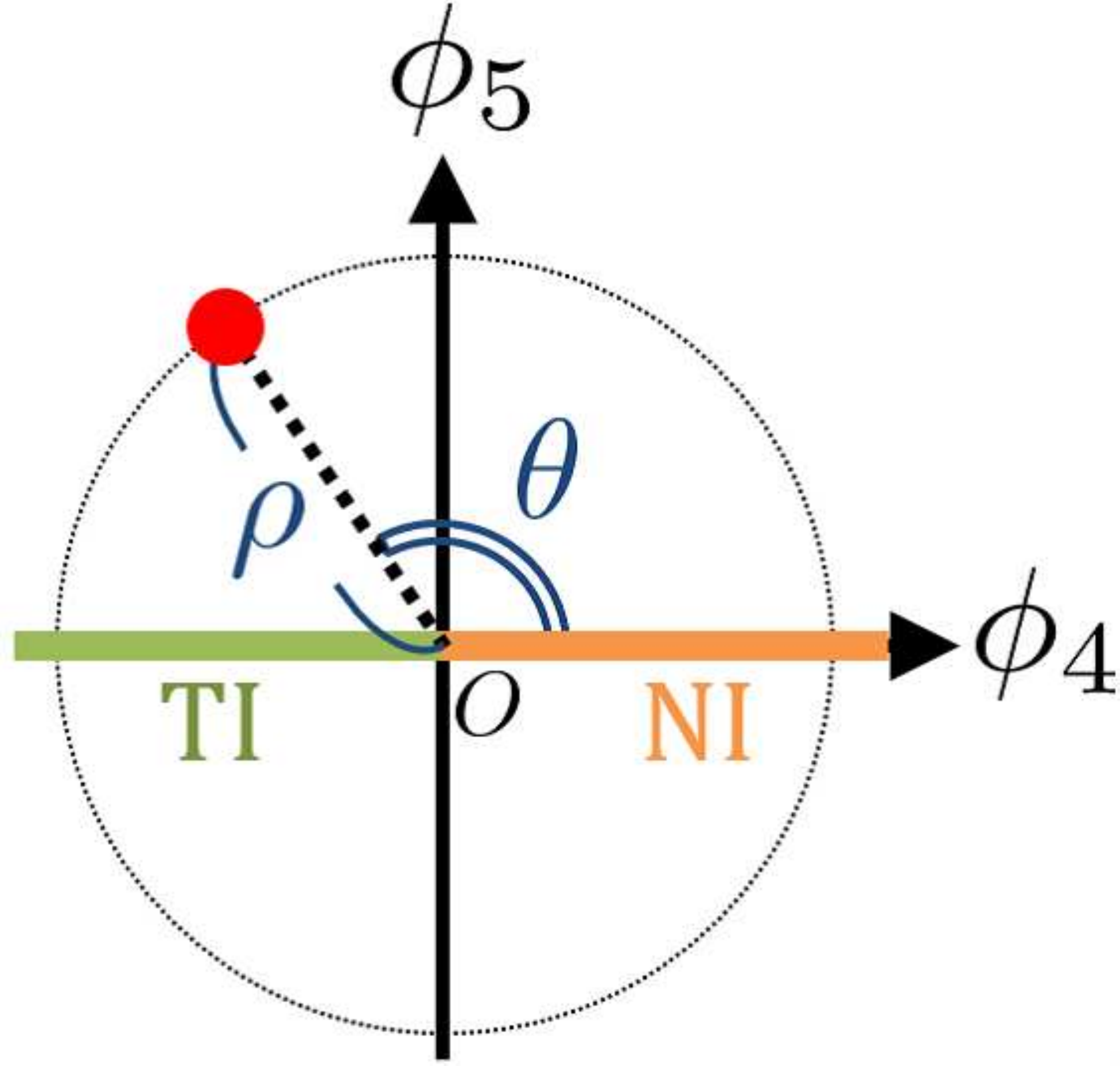}
\caption{
Relation between the axion field, $\theta$, and the non-magnetic and magnetic orders, $\phi_4$ and $\phi_5$.
In the model described by Eq.~\eqref{eq:H_TI}, $\phi_4$ corresponds to the order that characterizes the topological quantum phase transition,
and $\phi_5$ corresponds to the antiferromagnetic order.
In particular, a system preserving the time-reversal symmetry is located on the $\phi_4$ axis (i.e., $\phi_5=0$), which is classified as 
a normal insulator (NI) for $\phi_4>0$ or a topological insulator (TI) for $\phi_4<0$. 
}
\label{fig:1}
\end{center}
\end{figure}

From the above observation, we can expect that the non-magnetic order $\phi_4$ affects the critical electric field for the axion instability.
Figure~\ref{fig:1} illustrates the relation between the axion field, $\theta$, and the non-magnetic and magnetic orders, $\phi_4$ and $\phi_5$.
For a system with a fixed $\phi_4>0$ and $\phi_5=0$, a fluctuation in the magnetic order $\delta\phi_5$ induces a fluctuation of the axion field
$\delta \theta=\delta \phi_5/\phi_4$, which is inversely proportional to $\phi_4$.
Thus, $\theta$ is very sensitive to a small change of $\phi_5$ when $\phi_4$ is close to zero,
which makes the axion instability easily occur. 
This fact is already included in Eq.~\eqref{eq:Ecritical_OO}:
$E_0^{\rm crit}$ becomes smaller for smaller $g$ which corresponds to $\phi_4$ [see Eq.~\eqref{eq:theta-phi5}].
In the next section, we shall see that $\phi_4$ is related to the order that characterizes a topological phase transition.
A topological phase transition takes place at $\phi_4=0$ when $\phi_5=0$.

We further stress that as long as the system remains to be gapful and $\theta$ is well-defined, 
a large magnetic \revise{fluctuation}{order} of $\phi_5$ is not sufficient to obtain a large \revise{fluctuation of}{axion field} $\theta$.
In order to have a large $\theta$ \revise{fluctuation}{} that exceeds $2\pi$,
we also need to flip the sign of $\phi_4$ according to $\phi_4=\rho \cos\theta$.
This means that such a large \revise{fluctuation of}{} $\theta$ is most likely to \revise{occur}{arise}
near the topological phase transition point at $\phi_4=\phi_5=0$.
In the following discussions, we therefore assume that the system is near the topological phase transition point
and that $\phi_4$, as well as $\phi_5$, is a dynamical quantity which varies depending on applied electromagnetic fields.

\section{Low Energy Effective Theory of the Axion Electrodynamics}
\label{sec:lowenergy}

As discussed in Sec.~\ref{sec:review}, the non-magnetic order $\phi_4$ is also relevant to the axion instability.
In this section, taking into account $\phi_4$ as well as $\phi_5$,
we derive a low energy effective theory of the axion electrodynamics in insulating materials.

\subsection{Three-dimensional topological insulator as a platform of the axion field}
We start with the low-energy Hamiltonian of a non-interacting three-dimensional topological insulator:
\begin{eqnarray}
\mathcal{H}_{\rm TI}=\sum_{\bm k}\psi_{\bm
k}^{\dagger}H_{\rm TI}({\bm k})\psi_{\bm k},
\label{eq:H_TI}
\end{eqnarray}
where $\psi_{\bm k}^\dagger$ ($\psi_{\bm k}$) is a creation (an annihilation) operator of an electron with quasi-momentum $\bm k$ 
which has four (spin and orbital) internal degrees of freedom, and 
\begin{eqnarray}
H_{\rm TI}({\bm k})=\sum_{i=1,2,3} \hbar v_i k_i \gamma_i+\phi_4 \gamma_4. 
\label{eq:H_TI_k}
\end{eqnarray}
Here, $\gamma_{\mu=1,2,3,4}$ {are} the $4\times 4$ Hermitian gamma matrices that satisfy the
anti-commutation relation:\\
\begin{eqnarray}
\{\gamma_\mu, \gamma_\nu\}=2\delta_{\mu\nu}.
\end{eqnarray}
 $v_{i=1,2,3}$ is the electron velocity, and $\phi_4$ is the band-gap energy.
Equation~\eqref{eq:H_TI_k} is obtained as the $k\cdot p$ Hamiltonian of inversion symmetric topological insulators\cite{hzhang09}.
Indeed, by adding a higher-order regularization term $c{\bm k}^2 \gamma_4$ $(c>0)$ to $H_{\rm TI}(\bm k)$, 
one can confirm that the three dimensional ${\mathbb Z}_2$ topological number is given by the sign of $\phi_4$\cite{fukane07}:
When $\phi_4$ is negative (positive), the system is topologically non-trivial (trivial).

We note that $H_{\rm TI}({\bm k})$ has the time-reversal symmetry:
\begin{eqnarray}
T H_{\rm TI}({\bm k})
T^{-1}=H_{\rm TI} (-{\bm k}),
\end{eqnarray}
where $T$ is an antiunitary time-reversal operator which anti-commutes with $\gamma_{i=1,2,3}$ and commutes with $\gamma_4$.
For simplicity, we assume that $H_{\rm TI}({\bm k})$ also preserves the inversion symmetry:
\begin{eqnarray}
PH_{\rm TI}({\bm k})P^{-1}=H_{\rm TI}(-{\bm k}).
\end{eqnarray}
Namely, the inversion operator $P$
anti-commutes with $\gamma_{i=1,2,3}$ and commutes with $\gamma_4$.

From the symmetry properties of $H_{\rm TI}(\bm k)$ and $\gamma_4$, $\phi_4$ is invariant under both time-reversal and inversion.
Hence, $\phi_4$ in Eq.~\eqref{eq:H_TI_k} can be the non-magnetic order $\phi_4$ introduced in Sec.~\ref{sec:nonmag_fluctuation}.
On the other hand, there is no magnetic order ($\phi_5=0$), since the Hamiltonian preserves both the time-reversal and inversion symmetries.
The system described by the Hamiltonian \eqref{eq:H_TI_k} is located on the $\phi_4$ axis in Fig.~\ref{fig:1},
and the axion field can take either $0$ ($\phi_4>0$) or $\pi$ ($\phi_4<0$) mod $2\pi$, 
which corresponds to a normal insulator and a topological insulator, respectively.
The origin of the $\phi_4\phi_5$-plane ($\rho\equiv \sqrt{\phi_4^2+\phi_5^2}=0$) is
a quantum critical point of the topological phase transition between a normal insulator and a topological one.

In order to discuss axion instability, we need to take into account a term that breaks both the time-reversal and inversion symmetries.
The simplest form of such a term is $\phi_5\gamma_5$ with $\gamma_5\equiv \gamma_1\gamma_2\gamma_3\gamma_4$:
Under an operation $S=T$ or $P$, $\gamma_5$ is transformed to $S\gamma_5S^{-1}=-\gamma_5$, 
and hence $\phi_5$ is odd under both time-reversal and inversion.
This $\phi_5$ is the antiferromagnetic order $\phi_5$ introduced in Sec.~\ref{sec:nonmag_fluctuation}.
In the following discussions, we consider the minimal model that describes the axion instability, i.e., the Hamiltonian~\eqref{eq:H_TI}
with \revise{redefining}{replacing} $H_{\rm TI}(\bm k)$ \revise{as}{with}
\begin{eqnarray}
H({\bm k})=\sum_{i=1,2,3} \hbar v_i k_i \gamma_i+\phi_4 \gamma_4+\phi_5 \gamma_5. 
\label{eq:H_TI_k2}
\end{eqnarray}
Although $\phi_4$ and $\phi_5$ are material parameters, they should be determined self-consistently
in the presence of external electromagnetic fields so as to minimize the total energy including the electromagnetic ones
(see Sec.~\ref{sec:calculation}).

\subsection{Effective Lagrangian for the axion field}
\revise{In this section, we use the Fujikawa}{We use Fujikawa's} method\cite{fujikawa79,fujikawa80}
to relate $\phi_4$ and $\phi_5$ to the axion field $\theta$ and derive the effective Lagrangian.
In the Lagrangian formalism, the Hamiltonian~\eqref{eq:H_TI} with Eq.~\eqref{eq:H_TI_k2} is rewritten
in terms of the Lagrangian density \revise{}{as}
\begin{align}
\mathcal{L}_{\rm el}
=\bar{\psi}\left(i\hbar 
\sum_{\mu=0,1,2,3}v_{\mu}\Gamma_\mu 
D_\mu
-\phi_4\right)\psi
-i\phi_5\bar{\psi}{\Gamma_5}\psi,
\label{eq:Lel}
\end{align}
where $\psi({\bm x})=\sum_{\bm k}\psi_{\bm k}e^{i{\bm k}\cdot{\bm x}}$, 
$\Gamma_0=\gamma_4$, $\Gamma_{i=1,2,3}=\gamma_4\gamma_i$, 
$\Gamma_5=-i\Gamma_0\Gamma_1\Gamma_2\Gamma_3$, $v_0=1$ and 
$\bar{\psi}=\psi^{\dagger}\Gamma_0$.
Here we have replaced the partial derivative $\partial_{\mu}$ with the covariant derivative $D_{\mu}=\partial_{\mu}+ie A_{\mu}$
where $A_{\mu}$ is the gauge field.
In the absence of electromagnetic fields, an insulator has its pristine values of $\phi_4$ and $\phi_5$,
which are denoted by $\phi_4^{(0)}$ and $\phi_5^{(0)}$, respectively.
This means, there is an effective potential energy $\mathcal{V}_{\rm a}(\phi_4,\phi_5)$
which has a minimum at $(\phi_4^{(0)}, \phi_5^{(0)})$ in the $\phi_4\phi_5$-plane.
By taking into \revise{}{account} the contribution of this potential energy $\mathcal{V}_{\rm a}(\phi_4,\phi_5)$,
as well as the Lagrangian density $\mathcal{L}_{\rm em}$ of the electromagnetic fields,
the total Lagrangian density $\mathcal{L}$ is given by 
\begin{eqnarray}
\mathcal{L}=\mathcal{L}_{\rm em}+\mathcal{L}_{\rm el}-\mathcal{V}_{\rm a}(\phi_4,\phi_5),
\end{eqnarray}
from which, \revise{we obtain}{} the partition function is given in the path integral formalism as
\begin{eqnarray}
{\mathcal Z}=\int {\mathcal D}\psi{\mathcal D}\bar{\psi}
\exp\left[i\int d^4x \mathcal{L}\right].
\end{eqnarray}
We then perform the gauge transformation:
\begin{eqnarray}
\psi'=e^{-i\Gamma_5 \theta/2}\psi,
\quad
\bar{\psi'}=\bar{\psi}e^{-i\Gamma_5 \theta/2},
\end{eqnarray}
with $\theta$ defined by 
\begin{align}
\phi_4&=\rho\cos\theta,\\
\phi_5&=\rho\sin\theta,
\end{align}
so as to eliminate the time-reversal breaking term, $i\bar{\psi}\Gamma_5{\psi}$, in $\mathcal{L}_{\rm el}$.
This procedure, however, produces the $\theta$-term $\mathcal{L}_{\rm \theta}$ as the Jacobian of the path-integral measure, resulting in
\begin{eqnarray}
{\mathcal Z}&=&\int {\mathcal D}\psi'{\mathcal D}\bar{\psi'}
\exp\left[i\int d^4x 
\left(
\mathcal{L}'+\mathcal{L}_{{{\rm \theta}}}
\right)
\right],
\end{eqnarray}
where
\begin{eqnarray}
&&\mathcal{L}'=\mathcal{L}_{\rm em}+\bar{\psi'}\left(
i\hbar v_{\mu}\Gamma_{\mu}D'_{\mu}-\rho\right)\psi'
-\mathcal{V}_{\rm a}(\phi_4,\phi_5), 
\nonumber\\
&&D_{\mu}'=\partial_{\mu}+ie A_{\mu}+\frac{i}{2}\Gamma_5\partial_{\mu}\theta,
\end{eqnarray} 
and $\theta_0$ in $\mathcal{L}_{\rm \theta}$ is given by $\theta_0={\rm arg}\,(\phi_4^{(0)}+i\phi_5^{(0)})$.
By integrating with respect to $\psi'$ and $\bar{\psi}'$,
the following effective Lagrangian density ${\mathcal L}_{\rm b}$ for the bosonic fields $\phi_4$ and $\phi_5$
and the electromagnetic fields ${\bm E}$ and ${\bm B}$ is obtained:
\begin{eqnarray}
{\mathcal L}_{\rm b}&=&{\mathcal L}_{\rm em}+\mathcal{L}_{\rm \theta}-{\mathcal V}_{\rm a}(\phi_4,
\phi_5)
\nonumber\\
&&-i{\rm ln}\,{\rm det}\left[i\hbar \sum_{\mu=0,1,2,3} v_{\mu}\Gamma_{\mu}D'_{\mu}-\rho\right].
\end{eqnarray}
Here, the last term of the right hand side is the contribution from the integration of $\psi'$ and $\bar{\psi}'$,
which can be evaluated by using the derivative expansion in the low energy limit.
Since this term does not depend on $\theta$, it does not provide any correction of the $\theta$-term,
although it induces the kinetic term of $\theta$ and $\rho$, and the coupling between $A_{\mu}$ and the current $j_{\mu}$ of electron,
and \revise{renormalize}{renormalizes} $\epsilon$, $\mu$ and $\mathcal{V}_{\rm a}(\phi_4, \phi_5)$.
Thus, using the same notation of $\epsilon$, $\mu$ and $\mathcal{V}_{\rm a}(\phi_4,\phi_5)$ to represent the renormalized ones,
we eventually have the following low energy Lagrangian:
\begin{eqnarray}
{\mathcal L}_{\rm b}={\mathcal L}_{\rm em}+\mathcal{L}_{\rm \theta}-\mathcal{V}_{\rm a}(\phi_4, \phi_5),
\label{eq:effective_lagrangian}
\end{eqnarray} 
where we have omitted the kinetic term of $\theta$ and $\rho$ since we only discuss the static property in this paper.
Note that Eq. (\ref{eq:effective_lagrangian}) differs from the Lagrangian density introduced in Sec.~\ref{sec:OO}
in the potential energy of the axion field:
$\mathcal{V}_{\rm a}(\phi_4, \phi_5) $ in \revise{Eqs.}{Eq.}~\eqref{eq:effective_lagrangian}
is a function of $\phi_4$ as well as $\phi_5$ and can include higher-order terms of $\delta\theta$.

\subsection{Modified Maxwell equations and Hamiltonian density including the dynamical axion field}
By taking the functional derivative of Eq.~\eqref{eq:effective_lagrangian} with respect to $A_{\mu=0,1,2,3}$ and $\delta\theta$,
we obtain the modified Maxwell equations:
\begin{align}
&\nabla\cdot \bm B=0,\label{eq:nomonopole}\\
&\frac{1}{c}
\partial_t\bm B+
\nabla\times\bm E=\bm 0,\\
&\nabla\cdot {\bm D}=0,\label{eq:gausslaw}\\
-&\frac{1}{c}\partial_t{\bm D}
+
\nabla\times{\bm H}=\bm 0,\\
-&\frac{\alpha}{4\pi^2}\bm E\cdot\bm B
+\frac{\partial\mathcal V_{\rm a}}{\partial\delta\theta}
=0,\label{eq:axionlaw}
\end{align}
with the constitutive relations
\begin{align}
&\bm D=\epsilon\bm E+4\pi\bm P_\theta,\label{eq:ddefin}\\
&\bm H=\frac{1}{\mu}\bm B-4\pi\bm M_\theta,\label{eq:hdefin}\\
&\bm P_\theta=\frac{\alpha}{4\pi^2}\theta \bm B,\label{eq:ptheta}\\
&\bm M_\theta=\frac{\alpha}{4\pi^2}\theta \bm E.\label{eq:mtheta}
\end{align}
The Hamiltonian density corresponding to Eq.(\ref{eq:effective_lagrangian}) is given by
\begin{eqnarray}
{\mathcal H}_{\rm b}=\mathcal H_{\rm em}
+{\mathcal V}_{\rm a}(\phi_4,\phi_5)
\label{eq:hamiltonian}
\end{eqnarray}
where $\mathcal{H}_{\rm em}$ is the Hamiltonian density of the electromagnetic fields given by
\begin{eqnarray}
\mathcal H_{\rm em}
=
\frac{1}{8\pi}
\left( {\bm E}\cdot{\bm D}+{\bm B}\cdot{\bm H}
\right)
=
\frac{1}{8\pi}
\left(\epsilon {\bm E}^2+\frac{1}{\mu}{\bm B}^2
\right).
\label{eq:hamiltonianem}
\end{eqnarray}
It should be noted that the term \revise{}{that} corresponds to $\mathcal{L}_{\rm \theta}$ vanishes in the Hamiltonian formalism
because it is topological and does not contribute to the energy.

\section{Response to an Applied Electric Field}
\label{sec:calculation}

\subsection{Setup}

\begin{figure}[h]
\begin{center}
\includegraphics[width=6cm]{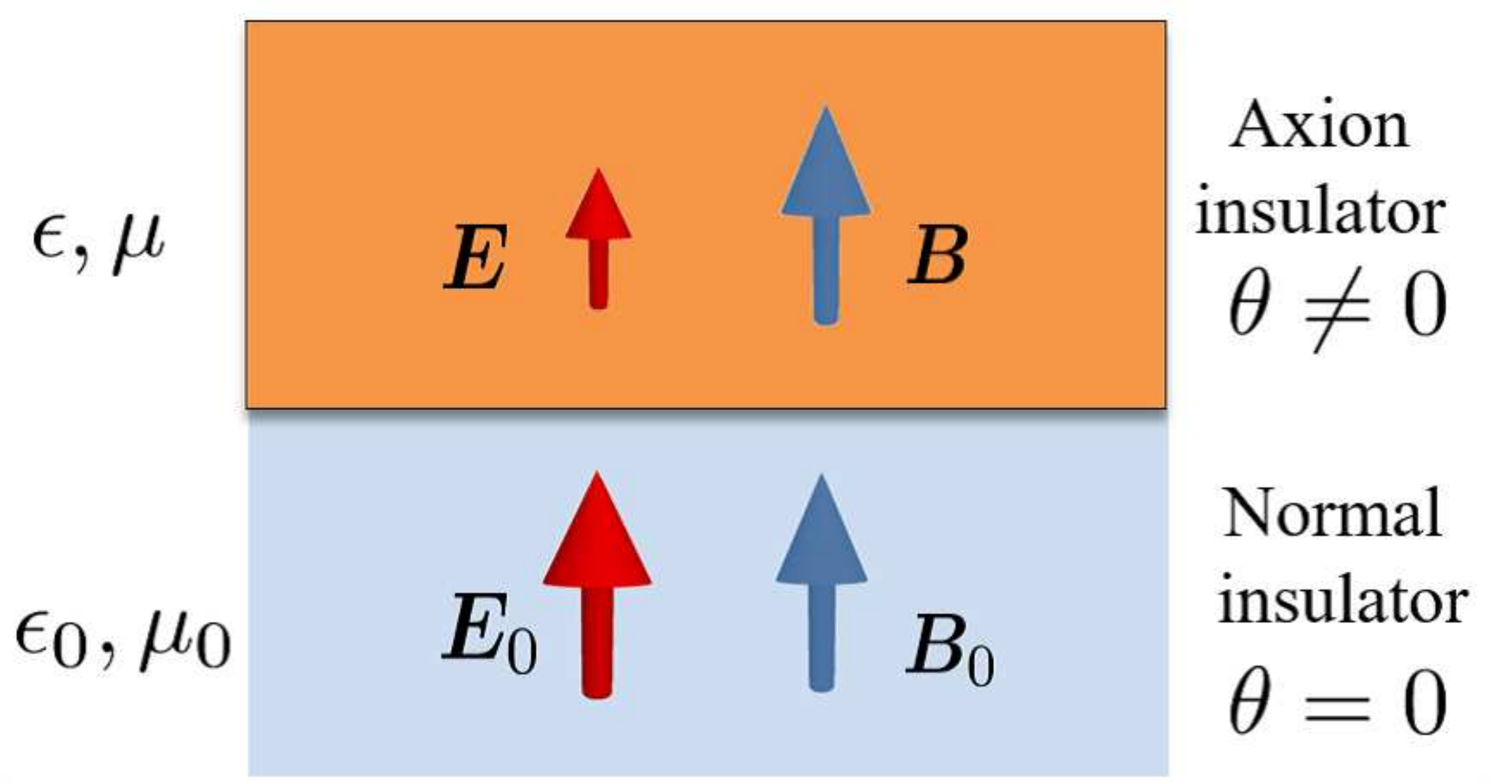}
\caption{Schematic of the setup.
We consider an interface between a normal insulator (bottom) and an axion insulator (top): 
The axion field is fixed to zero in the former,
whereas nonzero axion field can be induced 
\revise{due to the applied magnetic field in the latter}{in the latter in response to applied fields}.
Electric field $E_0$ is applied perpendicular to the interface. 
The electric field $E$ and the magnetic flux density $B$, as well as the axion field $\theta$, in the axion insulator
are determined so as to minimize the energy of the system.
The magnetic flux density in the normal insulator is the same as that in the axion insulator due to Eq.~\eqref{eq:nomonopole}.
}
\label{fig:setup}
\end{center}
\end{figure}

Following Ref.\cite{Ooguri12}, we consider an interface between two insulators described by Eq.~\eqref{eq:effective_lagrangian},
and apply an electric field $E_0$ perpendicular to the interface (Fig.~\ref{fig:setup}).
\revise{
We assume that both the insulators have $\theta_0=0$ but the potential energy $\mathcal{V}_a(\phi_4,\phi_5)$ for the bottom (top) one is steep (shallow)
so that the axion field is fixed to $\theta=0$ even in the presence of an applied electric field
(the nonzero axion field $\theta=\delta\theta$ is induced due to an applied electric field).
}
{
We assume that both the insulators have $\theta_0=0$.
We further assume that the potential energy $\mathcal{V}_{\rm a}(\phi_4,\phi_5)$ for the bottom (top) insulator is steep (shallow)
so that the axion field in the bottom insulator is fixed to $\theta=0$ even in the presence of applied fields
whereas a nonzero axion field $\theta=\delta\theta$ can be induced in the top insulator in response to applied fields.
}
In this paper, we refer to the bottom (top) insulator as a normal (an axion) insulator.
\revise{(}{}The normal insulator can be a vacuum.\revise{)}{}

Let $\epsilon_0$ and $\mu_0$ ($\epsilon$ and $\mu$) be the dielectric constant and the magnetic permeability of the normal (axion) insulator, respectively.
The boundary condition at the interface is obtained from Eqs.~\eqref{eq:gausslaw} and 
~\eqref{eq:nomonopole}
as
\begin{align}
&\epsilon E+\frac{\alpha}{\pi}\theta B=\epsilon_0 E_0, 
\label{eq:boundary}\\
&B=B_0\label{eq:boundaryhap}
\end{align}
where $E$ and $B(B_0)$ are the electric field and the magnetic flux density normal to the interface in the axion(normal) insulator, \revise{}{respectively}.

\subsection{Static \revise{Solutions}{solutions}}
In the following, we assume that energy dissipation of the system is large enough 
so that the system relaxes to a stationary state within a finite time after applying an electric field.
\revise{The static solution is}{Static solutions for the induced fields are} obtained by minimizing Eq.~(\ref{eq:hamiltonian})
as a function of $E$, $B$, $\rho$, and $ \theta$ under the boundary condition~\eqref{eq:boundary}.

First, we minimize the Hamiltonian density {\eqref{eq:hamiltonianem}} with respect to $E$ and $B$.
Substituting Eq.~(\ref{eq:boundary}) in Eq.~(\ref{eq:hamiltonianem}), we obtain
\begin{align}
{\mathcal H}_{\rm em}{(B,\theta)}
=\frac{\epsilon_{\rm eff}(\theta)}{8\pi\epsilon\mu}
\left(B-\tilde B(\theta)\right)^2
+\frac{(\epsilon_0E_0)^2}{8\pi \epsilon_{\rm eff}(\theta)},
\end{align}
with
\begin{align}
&\tilde B(\theta)=\frac{\theta/\Theta_0}{1+(\theta/\Theta_0)^2}\frac{\epsilon_0E_0}{\sqrt{\epsilon/\mu}},
\label{eq:def_Btheta}\\
&{\epsilon}_{\rm eff}(\theta)=\epsilon\left[1+\left({\theta}/{\Theta_0}\right)^2\right], \\
&\Theta_0=\frac{\pi}{\alpha}\sqrt{{\epsilon}/{\mu}}=4.3\times 10^{2}\sqrt{{\epsilon}/{\mu}}.
\label{eq:epsilontheta}
\end{align}
Hence, at the minimum of $\mathcal{H}_{\rm em}$, $B$ and $E$ are given as functions of $\theta$ as
\begin{align}
B&=\tilde B(\theta),
\label{eq:Btheta}\\
E&= \tilde E(\theta)
 \equiv \frac{\epsilon_0E_0}{ \epsilon_{\rm eff}(\theta)}, 
\label{eq:Etheta}
\end{align} 
where Eq.~\eqref{eq:Etheta} is obtained by substituting Eq.~\eqref{eq:Btheta} in Eq.~\eqref{eq:boundary}.
Equation~\eqref{eq:Etheta} indicates that $\epsilon_{\rm eff}(\theta)$ can be regarded as an effective dielectric constant modified by the axion field.

\begin{figure}[h]
\begin{center}
\includegraphics[width=7cm]{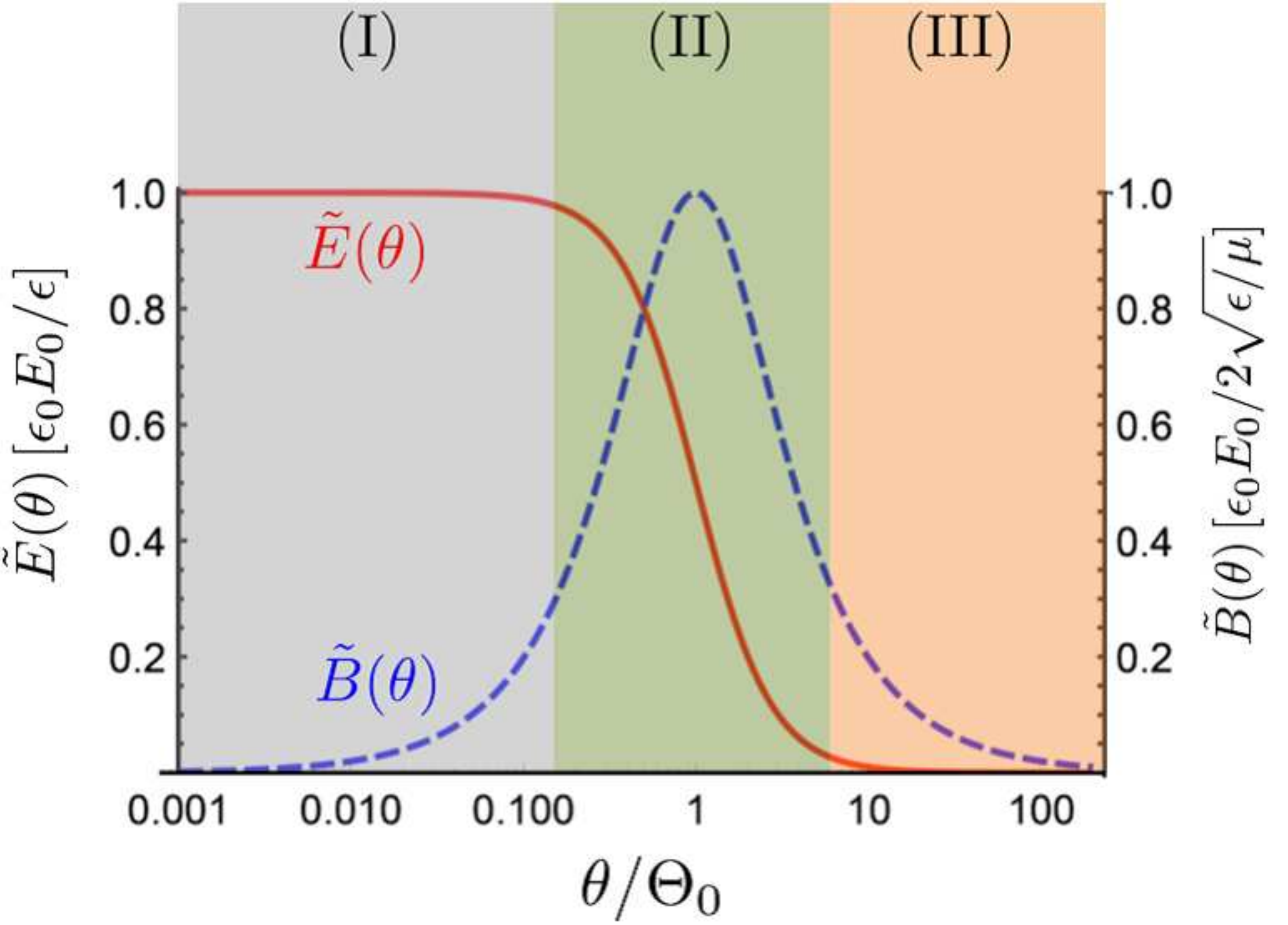}
\caption{
$\theta$ dependence of the stationary solutions for the electric field $\tilde{E}(\theta)$ and the magnetix flux density $\tilde{B}(\theta)$
\revise{inside the axion insulator}{} that minimize $\mathcal{H}_{\rm em}$ given \revise{in}{by} Eq.~\eqref{eq:hamiltonianem}.
\revise{Here $\tilde{E}(\theta)$, $\tilde{B}(\theta)$, and $\Theta_0$ are defined
in Eqs.~\eqref{eq:Etheta}, \eqref{eq:Btheta}, and \eqref{eq:epsilontheta}, respectively.}
{
Here, $\tilde{E}(\theta)$ [$\tilde{B}(\theta)$] is defined by Eq.~\eqref{eq:Etheta} [Eq.~\eqref{eq:def_Btheta}]
which is an even [odd] function of $\theta$.
$\theta$ is scaled by $\Theta_0$ defined in Eq.~\eqref{eq:epsilontheta}, and
}
$\tilde{E}(\theta)$ and $\tilde{B}(\theta)$ are scaled by \revise{the}{their} maximum values.
Note that the horizontal axis is logarithmic.
\revise{The models in Sec. \ref{sec:magnetic_fluctuation}, Sec. \ref{sec:ooguri-oshikawa} and Sec. \ref{sec:nonmag+mag} realize $\theta$ in the regions (I), (II) and (III), respectively.}
{The models in Secs.~\ref{sec:nonmag+mag} and \ref{sec:magnetic_fluctuation} realize $\theta$ in the regions (III) and (I), respectively.}
}
\label{fig:db}
\end{center}
\end{figure}

Figure~\ref{fig:db} illustrates the $\theta$ dependence of $\tilde{B}(\theta)$ and $\tilde{E}(\theta)$,
which shows that the nonzero axion field induces \revise{the}{a} magnetic flux density and screens the electric field instead.
This result can also be explained from the constituent equations~(\ref{eq:ddefin})-(\ref{eq:mtheta})
and the boundary condition (\ref{eq:boundary}) as follows:
 When \revise{the}{an} external electric field $ E_0 $ is applied,
\revise{the}{an} electric field $ E $ is generated to satisfy the boundary condition,
which induces $ M_ \theta $ via  Eq.~(\ref{eq:mtheta});
this $M_\theta$ works as a magnetic flux density $B$ and induces $P_\theta$ via Eq.~\eqref{eq:ptheta}
[indeed, from Eqs.~(\ref{eq:Btheta}) and (\ref{eq:Etheta}), $\tilde B(\theta)$ can be rewritten as
$\tilde B(\theta)=4\pi \mu M_\theta =\frac{\alpha}{\pi}\mu\theta \tilde E(\theta)$];
the induced $P_\theta$ screens a part of the applied electric field and increases the dielectric constant.
The solution that converged by repeating this process is Eqs. (\ref{eq:Btheta}) and (\ref{eq:Etheta}).

We note that $\tilde{E}(\theta)$ and $\tilde{B}(\theta)$ are functions of $\theta/\Theta_0$.
\revise{See Fig.(\ref{fig:db}).}{} 
Depending on $\revise{\theta}{|\theta|}/\Theta_0$, they have three different behaviors.
In region (I) \revise{($\theta\ll\Theta_0$)}{$|\theta|\ll\Theta_0$}, $\tilde{E}(\theta)$ keeps \revise{}{almost} a constant value $\epsilon_0E_0/\epsilon$,
and no significant magnetic field $\tilde{B}(\theta)$ is induced.
In region (II) \revise{($\theta\sim\Theta_0$)}{$|\theta|\sim\Theta_0$}, $\tilde{E}(\theta)$ begins to screened, and $\tilde{B}(\theta)$ is induced.
Finally, in region (III) \revise{($\theta\gg\Theta_0$)}{$|\theta|\gg\Theta_0$}, both $\tilde{E}(\theta)$ and $\tilde{B}(\theta)$ are screened.
When $|\theta|\gtrsim\Theta_0$, the contribution of $\mathcal L_{\rm \theta}$ becomes large compared to $\mathcal L_{\rm em}$,
and therefore, the interaction effect between the axion and electromagnetic fields becomes more significant.
Actually, as shown in Fig. \ref{fig:db},
$\tilde{E}(\theta)$ and $\tilde{B}(\theta)$ largely deviate from their values at $\theta=0$ around $|\theta|\sim \Theta_0$.
A typical value of $\Theta_0$ is in the order of $10^3 (\gg 2\pi)$, which means, a large axion field is required to observe the axion electromagnetism.
We also note that  $\tilde E(\theta)$ and $\tilde B(\theta)$ are not periodic in $\theta$ mod $2\pi$:
Although $2\pi$ periodicity in $\theta$ is imposed in a closed space-time with periodic boundary conditions,
this is not the case due to the existence of the interface, as pointed out in Ref.\cite{Ooguri12}.
\revise{In Fig.(\ref{fig:db})}{}

Next, we minimize the Hamiltonian density $\mathcal{H}_{\rm b}$ with respect to $\rho$ and $\theta$.
With the optimized $B$ and $E$ given by Eqs.~(\ref{eq:Btheta}) and (\ref{eq:Etheta}), respectively, 
the Hamiltonian density (\ref{eq:hamiltonian}) can be written as a function of $\rho$ and $\theta$ as
\begin{align}
\tilde{\mathcal H}_{\rm b}&=\tilde{\mathcal H}_{\rm em}(\theta)
+{\mathcal V_{\rm a}}(\phi_4, \phi_5),
\label{eq:hamiltoniantheta}\\
\tilde{\mathcal H}_{\rm em}(\theta)&=\frac{(\epsilon_0E_0)^2}{8\pi \epsilon \left[1+(\theta/\Theta_0)^2\right]},
\end{align}
with $\phi_4=\rho\cos\theta$ and $\phi_5=\rho\sin\theta$.
When both $\phi_4$ and $\phi_5$  fluctuate,
$\rho$ and $\theta$ can change independently, and stationary solutions are obtained by
solving $\partial \tilde{\mathcal H}_{\rm b}/\partial\rho=0$ and $\partial \tilde{\mathcal H}_{\rm b}/\partial \theta=0$.
The former equation reduces to $\partial{\mathcal V_{\rm a}}(\phi_4,\phi_5)/\partial \rho=0$, 
whose solution, denoted by $\tilde\rho(\theta)$, should be a periodic function of $\theta$ with the periodicity $2\pi$, i.e. $\tilde\rho(\theta+2\pi)=\tilde\rho(\theta)$, 
since $\mathcal{V}_{\rm a}(\phi_4,\phi_5)$ is in definition a periodic function of $\theta$.
Substituting $\tilde\rho(\theta)$ in Eq.~(\ref{eq:hamiltoniantheta}), we have
\begin{eqnarray}
\tilde
{\mathcal H}_{\rm b}(\theta)&=\tilde{\mathcal H}_{\rm em}(\theta)
+\tilde{\mathcal V_{\rm a}}(\theta),
\label{eq:Hb_theta0}
\end{eqnarray}
with
\begin{align}
\tilde{\mathcal V_{\rm a}}(\theta)
={\mathcal V_{\rm a}}(\tilde\rho(\theta)\cos\theta ,\tilde\rho(\theta)\sin\theta).
\end{align}
Therefore the induced axion field $\theta$ is determined by solving $\partial \tilde{\mathcal H}_{\rm b}(\theta)/\partial \theta=0$,
which reproduces Eq.~(\ref{eq:axionlaw}) with $E=\tilde{E}(\theta)$ and $B=\tilde{B}(\theta)$.

\subsection{Onset of the axion instability}
In Eq.~\eqref{eq:Hb_theta0}, the first term $\tilde{\mathcal H}_{\rm em}(\theta)$ has a maximum at $\theta=0$,
whereas the second term $\tilde{\mathcal{V}}_{\rm a}(\theta)$ has a minimum at $\theta=0$;
the former (latter) induces (suppresses) the axion instability.
 (Note that we have chosen $\theta_0=0$ in the axion insulator for simplicity. 
However, the calculation below is straightfoward even for the case of nonzero $\theta_0$.)
Since $\tilde{\mathcal H}_{\rm em}(\theta)$ is a monotonically decreasing function of $|\theta|$,
it provides a driving force for non-zero $\theta$.
Hence, even when $\theta$ is zero in the absence of the external electric field, a non-zero $\theta$ can be realized
if $E_0$ exceeds a critical value $E_0^{\rm crit}$ and the first term is
dominant in $\tilde{\mathcal H}_{\rm b}(\theta)$. 
Below, we evaluate $E_0^{\rm crit}$.

We define the effective square mass of axion  as \revise{as}{} a curvature of $\tilde{\mathcal H}_{\rm b}(\theta)$ at $\theta=0$:
\begin{align}
M_{\rm eff}^2
=\left.\frac{\partial^2\tilde{\mathcal H}_{\rm b}(\theta)}{\partial\theta^2}
\right|_{\theta=0}
=M_{\rm em}^2+M_{\rm a}^2
\end{align}
with
\begin{align}
&M_{\rm em}^2
=\left.\frac{\partial^2\tilde{\mathcal H}_{\rm em}(\theta)}{\partial\theta^2}
\right|_{\theta=0}
=-\frac{1}{\Theta_0^2}\frac{(\epsilon_0E_0)^2}{4\pi\epsilon}{\le0}
,\\
&M_{\rm a}^2=\left.\frac{\partial^2{\tilde{\mathcal V}_{\rm a}}(\theta)}{\partial\theta^2}
\right|_{\theta=0}{>0}.
\label{eq:mass}
\end{align}
Here, the sign of $M_{\rm a}^2$ is fixed to be positive by definition of $\mathcal{V}_{\rm a}(\phi_4,\phi_5)$.
Note that sign change of the squared mass of the axion $ M_{\rm eff}^2 $
occurs when the external electric field $ E_0 $ is bigger than a certain threshold.
Defining the threshold $E_0^{\rm crit}$ as $E_0$ \revise{which}{that} satisfies $M_{\rm eff}^2=0$, we obtain
\begin{align}
E_0^{\rm crit}=\frac{\sqrt{4\pi\epsilon}\Theta_0 M_{\rm a}}{\epsilon_0}.
\label{eq:Ecrit}
\end{align}
When $E_0<E_0^{\rm crit}$, $\tilde{\mathcal H}_{\rm b}(\theta)$ has positive curvature $M_{\rm eff}^2>0$ (bradyonic),
so $\theta=0$ remains to be a (at least local) minimum of $\tilde{\mathcal H}_{\rm b}(\theta)$.
On the other hand when $E_0>E_0^{\rm crit}$, $\tilde{\mathcal H}_{\rm b}(\theta)$ has negative curvature $M_{\rm eff}^2<0$ (tachyonic),
so $\theta=0$ becomes unstable, i.e., the axion instability occurs. 

Although one may think that $E_0^{\rm crit}$ is too large to cause the axion instability in realistic systems,
Eq.~(\ref{eq:Ecrit}) indicates that $E_0^{\rm crit}$ takes a lower value {for smaller $M_{\rm a}$}.
Note that \revise{Eqs.}{Eq.}~(\ref{eq:mass}) is rewritten as
\begin{align}
M_{\rm a}^2={\tilde\rho(0)^2}M_5^2
\label{eq:mass2},
\end{align}
with
\begin{align}
M_5^2=\left.\frac{\partial^2{\tilde{\mathcal V}_{\rm a}}{(\theta)}
}{\partial\phi_5^2}
\right|_{\phi_5=0}.
\label{eq:massm5}
\end{align}
Therefore, there are two ways to reduce the critical value $E_0^{\rm crit}$:
One is to reduce the value of $M_5$ by going near the quantum critical point of the antiferromagnetic order, as discussed in Ref.\cite{Ooguri12};
the other is to reduce the value of $\tilde\rho(0)$ by going near the topological quantum phase transition {as discussed in the previous section.

\subsection{Correspondence with Ooguri-Oshikawa's theory}
For a small fluctuation of $\theta\simeq\delta\theta$,
the potential term in Eq.~\eqref{eq:Hb_theta0} can be approximated up to the second order in $\delta\theta$:
\begin{eqnarray}
\tilde{\mathcal V}_{\rm a}(\theta)=\frac{M_{\rm a}^2}{2}(\delta\theta)^2=\frac{M_5^2}{2}(\delta \phi_5)^2.
\label{eq:potentialap}
\end{eqnarray}
Comparing this equation with Eqs.~\eqref{eq:axion} and \eqref{eq:theta-phi5},
we obtain the correspondence relation: $\rho(0)\leftrightarrow g$ and $M_{\rm a}^2/2 \leftrightarrow g^2 Jm^2$.
Therefore, $E_0^{\rm crit}$ in Eq.~(\ref{eq:Ecrit}) coincides with that in Eq.~(\ref{eq:Ecritical_OO}).

\section{Model analysis}
\label{sec:model}
In this section, we analyze the axion instability based on  model potentials.
Before going to the detailed analysis, we first present a general consideration.
In the initial state, the system is an ordinary (non-topological) insulator with time-reversal invariance.  
It should be noted that $\theta=0$ is not the global minimum of $\tilde{\mathcal H}_{\rm b}(\theta)$ for $E_0\neq 0$: 
because the first term of Eq.~(\ref{eq:Hb_theta0}) is a decreasing function of $|\theta|$ and the second term is periodic in $\theta$,
the inequality $\tilde{\mathcal H}_{\rm b}(\theta) > \tilde{\mathcal H}_{\rm b}(\theta+{\rm sgn}(\theta) 2\pi)$ always holds.
Therefore, once the the axion instability occurs, 
\revise{
a large fluctuation of $\theta$ such that $|\theta|/{\Theta_0}\gg 1$ is expected,
at which the first term of Eq.~\eqref{eq:Hb_theta0} is suppressed.
Furthermore, Eqs.~(\ref{eq:Btheta}) and (\ref{eq:Etheta}) imply that such a large fluctuation of $\theta$ 
induces the magnetic field $B$ and screens the electric field $E$ inside the axion insulator. 
}{
a large axion field $\theta$ such that $|\theta|/{\Theta_0}\gg 1$ is expected to emerge,
unless the potential $\tilde{\mathcal V}_{\rm a}(\theta)$ has a singularity.
Note that $\tilde{\mathcal H}_{\rm em}(\theta)$ scales as a function of $\theta/\Theta_0$.
Furthermore, Fig.~\ref{fig:db} indicates that such a large $\theta$ leads to almost complete screening of the electric field
and induction of a small magnetic field inside the axion insulator.
On the other hand when $\tilde{\mathcal V}_{\rm a}(\theta)$ diverges at a certain $\theta$, 
the induced $|\theta|$ is bounded to be less than $2\pi\ll \Theta_0$. 
In this case, screening of $E$ and induction of $B$ are small as seen from Fig.~\ref{fig:db}.
}
Hence, although the critical electric field $E_0^{\rm crit}$, 
which is determined by behaviors around $\theta=0$, \revise{reproduces}{agrees with} the result by Ooguri and Oshikawa\cite{Ooguri12},
we find that the resulting behaviors of $E$ and $B$ are totally different.
Below, we consider two model potentials \revise{}{with and without non-magnetic fluctuations which correspond to
analytic and singular $\tilde{\mathcal V}_{\rm a}(\theta)$, respectively.}

\subsection{Instability due to coexisting non-magnetic and magnetic fluctuations}
\label{sec:nonmag+mag}

First, we consider a model hosting both non-magnetic and magnetic fluctuations. 
As we will see below, this model corresponds to the region (III) in Fig.\ref{fig:db}:
The coexistence of non-magnetic and magnetic fluctuations makes it possible to induce a huge value of $\theta$,
but it suppresses $B$ and $E$ according to Eqs.(\ref{eq:Btheta}) and (\ref{eq:Etheta}).

As an example of a $2\pi$ periodic potential, consider the following model potential: 
\begin{eqnarray}
\tilde{\mathcal V}_{\rm a}(\theta)={M_{\rm a}^2}(1-\cos\theta), 
\label{eq:potential2}
\end{eqnarray}
which has the minimum value $0$ at $\theta={2\pi}{n}$ and 
the maximum value $2M_{\rm a}^2$ at $\theta={2\pi}({n}+1/2)$ with $n$ being an integer.
\revise{Since Eq.~\eqref{eq:potential2} satisfies Eq.~\eqref{eq:mass}, the critical electric field $E_0^{\rm crit}$ is given by Eq.~\eqref{eq:Ecrit}.}{}
In the presence of the external electric field $E_0$,
the $\theta$ dependence of $\tilde{\mathcal H}_{\rm b}(\theta)$ is given by
\begin{align} 
&\tilde{\cal H}_{\rm b}(\theta)=\tilde{\cal H}_{{\rm em}}(0)\nonumber\\
&+\frac{M_{\rm a}^2}{2}
\left[-\frac{\theta^2}{1+(\theta/\Theta_0)^2}
\left( \frac{E_0}{E_0^{\rm crit}}\right)^2+{2}(1-\cos\theta)
\right].
\label{eq:Hb_theta-Hb_0}
\end{align}
In Fig.\ref{fig:h_2}, we plot \revise{the left-side of Eq.~\eqref{eq:Hb_theta-Hb_0} for various $E_0$}
{$\tilde{\mathcal H}_{\rm b}(\theta)-\tilde{\mathcal H}_{\rm b}(0)$ for $E_0/E_0^{\rm crit}=0.9, 1.0$ and $1.1$}
as a function of $\theta$.
\revise{}{Here, we have assumed $\epsilon=\mu=1$ and used $\Theta_0=4.3\times 10^2$ [see Eq.~\eqref{eq:epsilontheta}].}
In the scale of $\theta$ shown in Fig.~\ref{fig:h_2}(a), $\theta=0$ seems to be a maximum of $\tilde{\mathcal H}_{\rm b}(\theta)$, 
but actually this point is a local minimum (maximum) for $E_0<E_0^{\rm crit}$ ($E_0>E_0^{\rm crit}$) as seen in Fig.~\ref{fig:h_2}(b).
Because of the steep peak shown in Fig.~\ref{fig:h_2}(a),
which comes from the first term \revise{of the right-hand side of}{in the square bracket in} Eq.~\eqref{eq:Hb_theta-Hb_0},
when $E_0$ exceeds the critical value $E_0^{\rm crit}$ and the axion instability occurs, 
$|\theta|$ becomes much larger than $\Theta_0$ as we mentioned \revise{in the previous section}{above}.
The system is \revise{indeed}{} expected to end up with the first stationary point $\theta_{\rm min}$,
\revise{}{which is} estimated \revise{below}{as follows}.

From $\left.\partial \tilde{\mathcal H}_{\rm b}(\theta)/\partial \theta\right|_{\theta=\theta_{\rm min}}=0$,
we have
\begin{eqnarray}
\Theta_0\left(\frac{E_0}{E^{\rm crit}_0}\right)^2
f(\theta_{\rm min})=
\sin \theta_{\rm min},
\label{eq:1111}
\end{eqnarray}
with
\begin{eqnarray}
f(\theta_{\rm min})=\frac{
(\theta_{\rm min}/\Theta_0)}{\left[1+(\theta_{\rm min}/\Theta_0)^2\right]^2}.
\end{eqnarray}
Since $\Theta_0$ in Eq.~(\ref{eq:1111}) is much larger than the right-hand side in Eq.~(\ref{eq:1111})
[a typical value of $\Theta_0$ is in the order of $10^3$, see Eq.~\eqref{eq:epsilontheta}],
$f(\theta_{\rm min})$  should be much smaller than 1,
\revise{}{which means $|\theta_{\rm min}|/\Theta_0\gg 1$ and $f(\theta_{\rm min})\simeq (\theta_{\rm min}/\Theta_0)^{-3}$.}
\revise{Therefore, for $\theta_{\rm min}\ne0$, we obtain $\theta_{\rm min}/\Theta_0\gg1$, where Eq.(\ref{eq:1111}) reduces to}
{The first positive solution of Eq.~\eqref{eq:1111} arises around where the left-hand side of Eq~\eqref{eq:1111} decreases to unity, i.e.,}
\revise{Here the $\pm$ branches come from time-reversal symmetry.
Because $|\sin\theta_{\rm min}|\sim O(1)$, $\theta_{\rm min}$ is estimated as}{}
\begin{eqnarray}
\revise{\theta_{\rm min}\simeq
\pm
\left[\Theta_0^4
\left(\frac{E_0}{E^{\rm crit}_0}\right)^2\right]^\frac{1}{3}.}
{\theta_{\rm min}\simeq \Theta_0^{4/3}\left(\frac{E_0}{E^{\rm crit}_0}\right)^{2/3}.}
\label{eq:theta_model2}
\end{eqnarray}   

\revise{For $\theta_{\rm min}>0$, from Eqs.(\ref{eq:Btheta}) and (\ref{eq:Etheta}), we can estimate $B$
and $E$ inside the axion insulator as }
{With this $\theta_{\rm min}$, the electromagnetic fields $B$ and $E$ inside the axion insulator are
evaluated from Eqs.~\eqref{eq:Btheta} and \eqref{eq:Etheta} as}
\begin{align}
B&=\tilde B(\theta_{\rm min})
\simeq \Theta_0^{-{\frac{1}{3}}}\frac{\epsilon_0 E_0^{\rm crit}}{\sqrt{\epsilon/\mu}}\left(\frac{E_0}{E^{\rm crit}_0}\right)^{\frac{1}{3}},
\label{eq:B_model2}
\\
E&=\tilde E(\theta_{\rm min})
\simeq \Theta_0^{-\frac{2}{3}}\frac{\epsilon_0E_0^{\rm crit}}{\epsilon }\left(\frac{E_0}{E^{\rm crit}_0}\right)^{-\frac{1}{3}}.
\label{eq:E_model2}
\end{align}
\revise{}{In Fig.~\ref{fig:dbt2}, we plot the axion field $\theta$ and the electromagnetic feilds $E$ and $B$
induced in the axion insulator as functions of an applied electric field $E_0$.
One can see that $\theta$ suddenly increases to a huge value at $E_0=E_0^{\rm crit}$, which almost completely screens the electric field $E$.
The axion field also induces a magnetic flux density $B$,
but its amplitude is quite small since the induced $\theta$ is in the region (III) of Fig.~\ref{fig:db}.
For comparison, the results of Ooguri-Oshikawa's theory are shown in Fig.~\ref{fig:dbtoo}.}

\revise{}{Since $\tilde{\mathcal H}_{\rm b}(\theta)$ is an even function of $\theta$, $\theta=-\theta_{\rm min}$ is also a solution of Eq.~\eqref{eq:1111}.
The sign of the axion field is spontaneously determined. In other words, time-reversal symmetry is spontaneously broken at the onset of the instability.
Accordingly, the direction of the induced magnetic field is determined.}

\revise{
When $E_0>E_0^{\rm crit}$, $\tilde{\mathcal H}_{\rm b}(\theta)$ has a lot of minima with
nonzero $\theta_{\rm min}$, and thus time-reversal symmetry is spontaneously
broken by developing the antiferromagnetic order $\phi_5$.
which clearly shows that the electric field $E$ is almost completely screened.
}{}

\begin{figure}[h]
\begin{center}
\includegraphics[width=8cm]{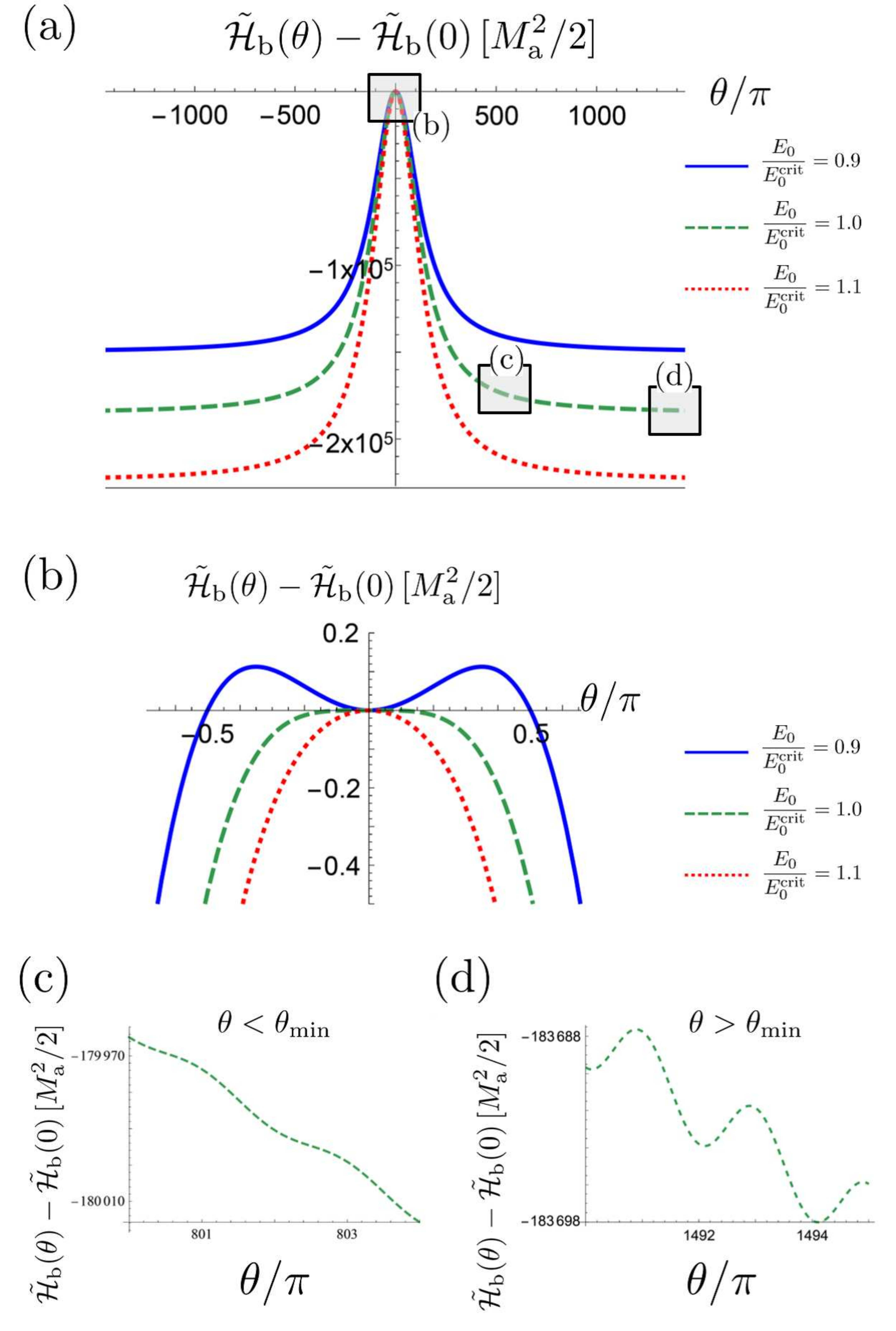}
\caption{
\revise{$\tilde{\cal H}_{\rm b}(\theta)$ in the model with coexisting non-magnetic and magnetic fluctuations.}
{
Hamiltonian density $\tilde{\cal H}_{\rm b}(\theta)$ [Eq.~\eqref{eq:Hb_theta-Hb_0}]
in the model with coexisting non-magnetic and magnetic fluctuations 
as a function of the induced axion field $\theta$.
We assume an axion insulator with $\epsilon=\mu=1$ and use $\Theta_0=4.3\times 10^2$.
(a) shows behavior of $\tilde{\cal H}_{\rm b}(\theta)$ in a wide range of $\theta$,
and (b)--(d) are the enlarged views in the regions marked by boxes in (a).
(b) The point $\theta=0$ changes from a local minimum to a maximum at $E_0=E_0^{\rm crit}$.
Due to the $\cos\theta$ term in Eq.~\eqref{eq:Hb_theta-Hb_0}, 
(c) $\tilde{\mathcal H}_{\rm b}(\theta)$ exhibits a wavy curve for $\theta<\theta_{\rm min}$, and
(d) local minima periodically appear for $\theta>\theta_{\rm min}$,
where $\theta_{\rm min}$ is the $\theta$ for the first local minimum at $\theta>0$.
We assume that the system under $E_0>E_0^{\rm crit}$ relax to the first local minumum, $\theta=\theta_{\rm min}$, 
and derive the electromagnetic fields [Eqs.~\eqref{eq:B_model2} and \eqref{eq:E_model2}] inside the axion insulator.
}\\
}
\label{fig:h_2}
\end{center}
\end{figure}

\begin{figure}[h]
\begin{center}
\includegraphics[width=7cm]{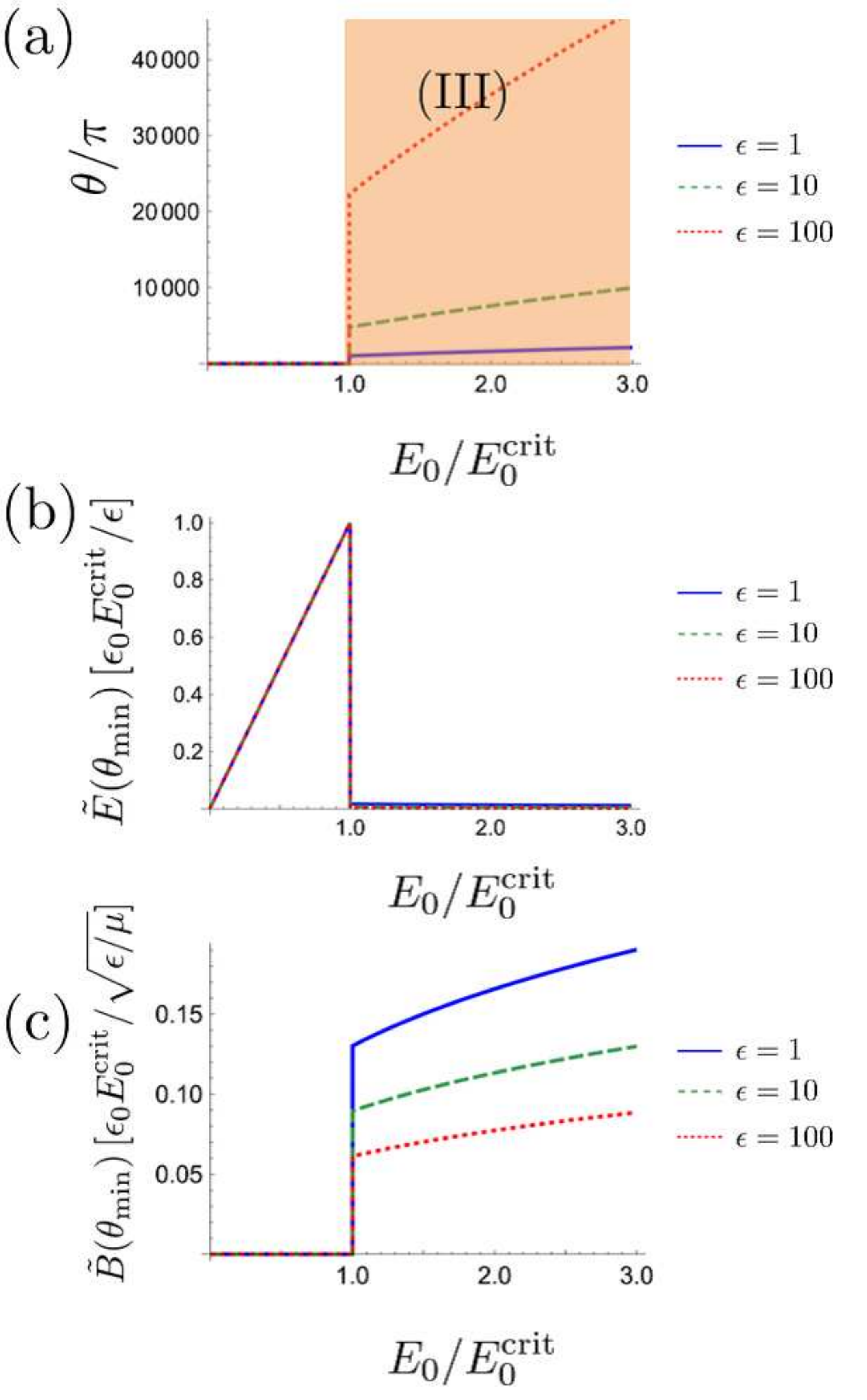}
\caption{
\revise{}{Behaviors of (a) the axion field $\theta$, (b) the electric field $E$, and (c) the magnetic flux density $B$
inside an axion insulator in response to an applied electric field $E_0$ in the model with coexisting non-magnetic and magnetic fluctuations,
where the values at $E_0>E_0^{\rm crit}$ are given by Eqs.~\eqref{eq:theta_model2}, \eqref{eq:E_model2} and \eqref{eq:B_model2}, respectively.
We choose $\mu=1$ and $\epsilon=1, 10$ and $100$.
From Eq.~\eqref{eq:epsilontheta}, the corresonding values of $\Theta_0$ are $4.3\times 10^2, 1.4\times 10^3$ and $4.3\times 10^2$, respectively.
Above the critical field, a huge axion field is induced[region (III) of Fig. 3], which almost completely screens the electric field.
A magnetic flux density is induced at $E_0>E_0^{\rm crit}$ but its amplitude is small.
}\\
}
\label{fig:dbt2}
\end{center}
\end{figure}

\subsection{Instability due to magnetic fluctuations}
\label{sec:magnetic_fluctuation}

In this subsection, 
we consider a model with fixed {$\phi_4=\tilde\rho(0)>0$}, where only the antiferromagnetic order $\phi_5$ can fluctuate.   
As is shown below,  only a small $\theta$ can be induced in this case, which corresponds to the region (I) \revise{of}{in} Fig.\ref{fig:db}. 

Since $\phi_4=\rho(\theta)\cos\theta$ is a constant $\rho(0)$, we have $\phi_5=\tilde\rho(\theta)\sin\theta=\tilde\rho(0)\tan\theta$.
Then, we assume the following quadratic potential for $\phi_5$,
\begin{eqnarray}
\tilde{\mathcal V}_{\rm a}(\theta)=\frac{M_5^2}{2}\phi_5^2 
{=\frac{M_{\rm a}^2}{2}\tan^2\theta}, 
\label{eq:potential1}
\end{eqnarray}
which has \revise{the}{a} minimum at $\theta=0$.
\revise{and satisfies Eqs.(\ref{eq:mass}) and (\ref{eq:massm5}), 
so the critical electric field $E_0^{\rm crit}$ is given by Eqs.(\ref{eq:Ecrit}) and (\ref{eq:mass2}).}{}
We note that the induced $|\theta|$ is $\pi/2$ at most because there is an infinitely high potential barrier at $|\theta|=\pi/2$.

\begin{figure}[h]
\begin{center}
\includegraphics[width=8cm]{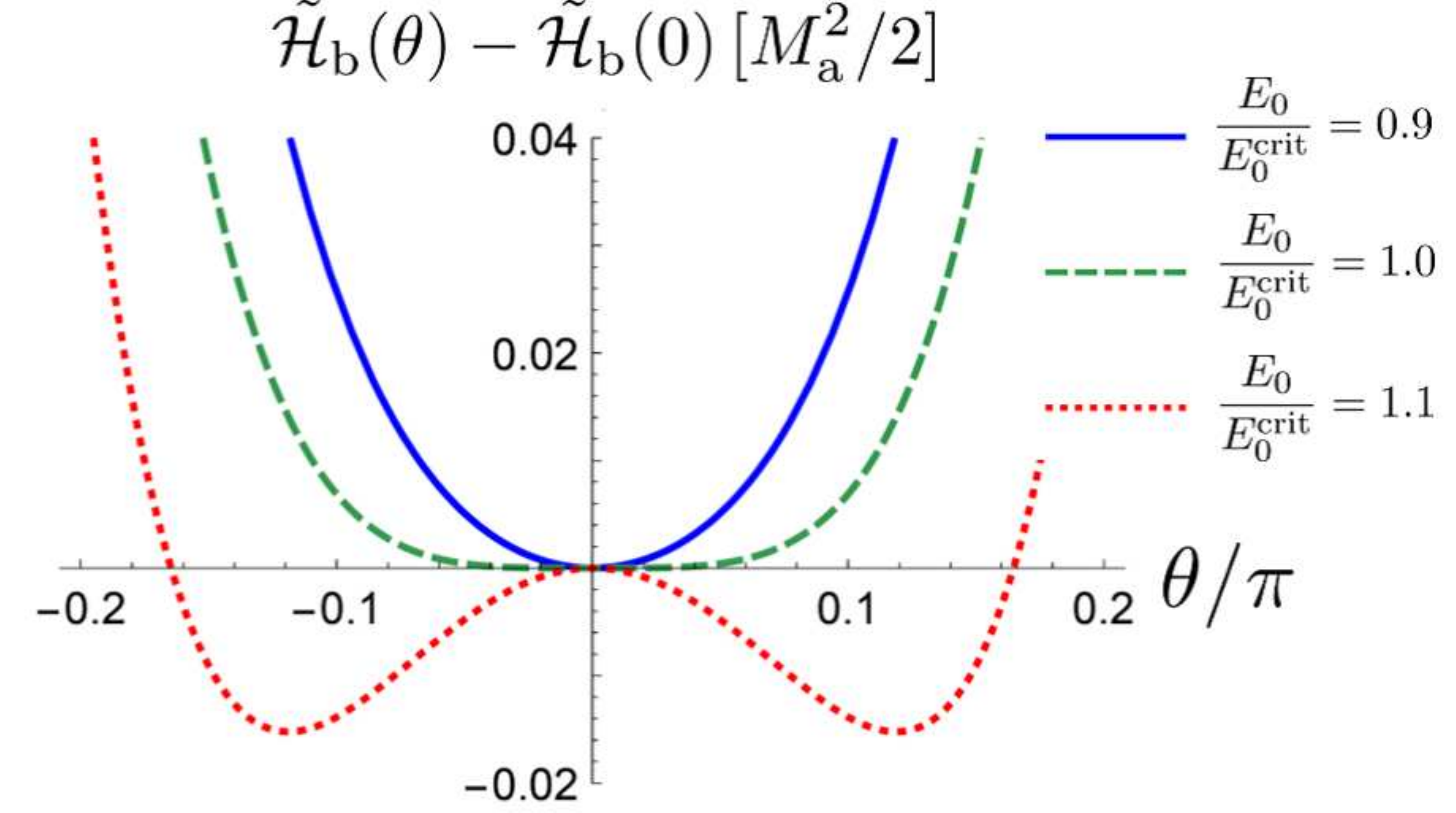}
\caption{\revise{${\mathcal H}_{\rm b}(\theta)$ in the model with magnetic fluctuations.}
{
Hamiltonian density ${\mathcal H}_{\rm b}(\theta)$ [Eq.~\eqref{eq:Hb_model1}] in the model only with magnetic fluctuations.
The point $\theta=0$ changes from a minimum to a local maximum at $E_0=E_0^{\rm crit}$ and double minima arise for $E_0>E_0^{\rm crit}$.
Because the potential~\eqref{eq:potential1} diverges at $\theta=\pm\pi/2$, the position of the minima is restricted in $|\theta|<\pi/2$ and 
goes to $|\theta|=\pi/2$ as $E_0/E_0^{\rm crit}\to \infty$.}
}
\label{fig:h_1}
\end{center}
\end{figure}

\revise{The energy density is given by $\tilde{\mathcal H}_{\rm b}(\theta)$ in Eq.(\ref{eq:hamiltonian})
When $E_0>E_0^{\rm crit}$, we expect the induced $|\theta|\ll \Theta_0$, we have }
The energy density is given by $\tilde{\mathcal H}_{\rm b}(\theta)$ in Eq.~\eqref{eq:Hb_theta0}
with the potential~\eqref{eq:potential1}.
Since $|\theta|\ll\Theta_0$ even at $E_0>E_0^{\rm crit}$, $\tilde{\mathcal H}_{\rm b}(\theta)$ is approximately given by
\begin{align}
\revise{}
{
\tilde{\cal H}_{\rm b}(\theta)\simeq\tilde{\cal H}_{\rm em}(0)+\frac{M_{\rm a}^2}{2}
\left[-\theta^2\left( \frac{E_0}{E_0^{\rm crit}}\right)^2+\tan^2\theta\right].
}
\label{eq:Hb_model1}
\end{align}
In Fig.~\ref{fig:h_1}, we show $\tilde{\mathcal H}_{\rm b}(\theta)-\tilde{\mathcal H}_{\rm b}(0)$
for \revise{various $E_0$}{$E_0/E_0^{\rm crit}=0.9, 1.0$ and $1.1$}
as a function of $\theta$.
When $E_0>E_0^{\rm crit}$, $\tilde{\mathcal H}_{\rm b}(\theta)$ has double minima \revise{with}{at} nonzero $\theta$,
and thus time-reversal symmetry is spontaneously broken by 
\revise{developing}{choosing one of the two minima, resulting in}
the antiferromagnetic order $\phi_5$.

\begin{figure}[h]
\begin{center}
\includegraphics[width=7cm]{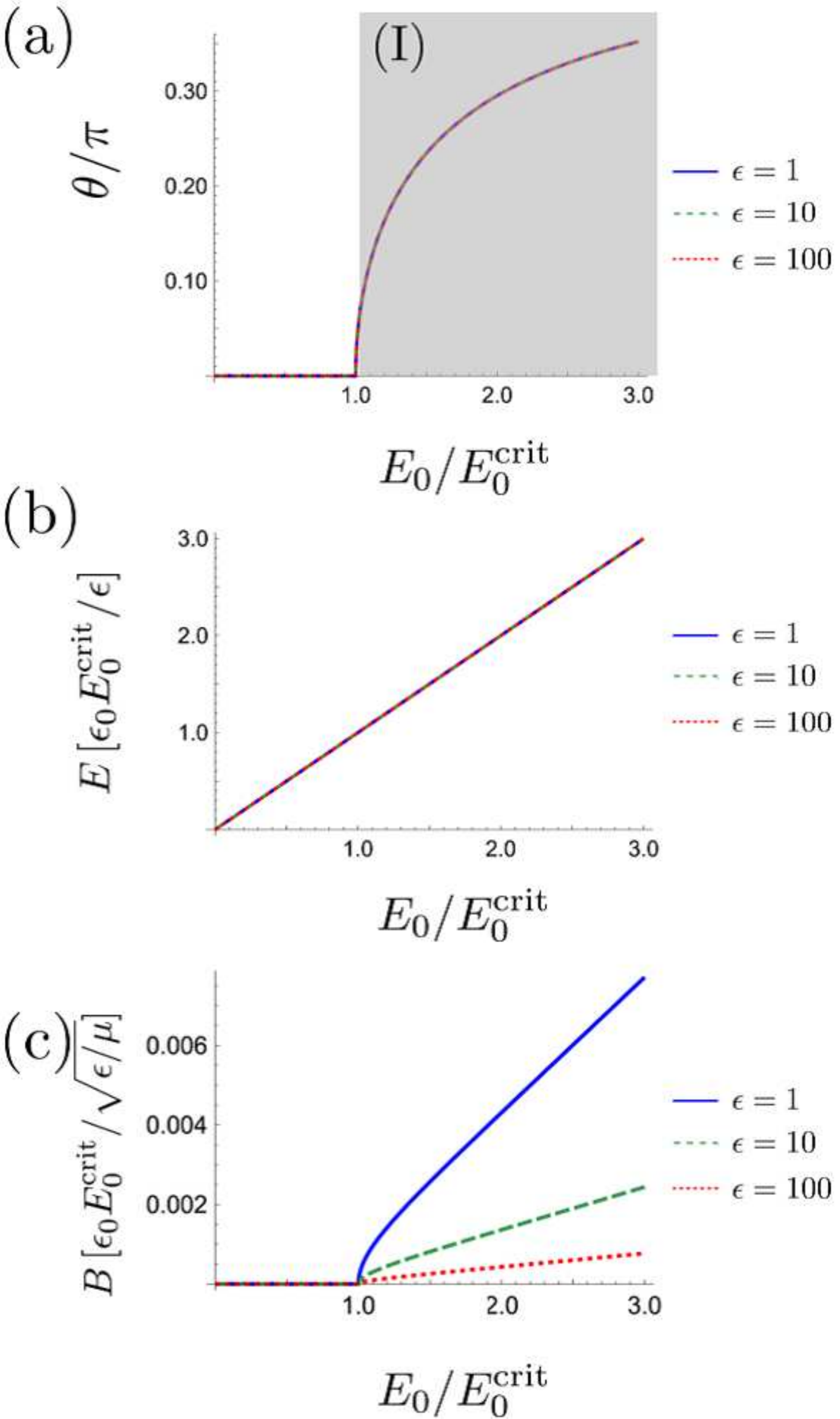}
\caption{
\revise{Applied electric field dependence of 
$\theta$, $E$ and $B$ with magnetic  fluctuation. 
(a)$\theta/\pi$ is plotted as a function of $E_{0}$. 
(b)$\theta/\Theta_{0}$ is plotted as a function of $E_{0}$. 
(c)$E$ is plotted as a function of $E_{0}$. 
(d)$B$ is plotted as a function of $E_{0}$.}
{
Behaviors of (a) the axion field $\theta$, (b) the electric field $E$, and (c) the magnetic flux density $B$
inside an axion insulator in response to an applied electric field $E_0$ in the model only with magnetic fluctuations,
where $\theta$ at $E_0>E_0^{\rm crit}$ is obtained by numerically finding the minima of Eq.~\eqref{eq:Hb_model1},
from which $E$ and $B$ are calculated using Eqs.~\eqref{eq:Etheta} and \eqref{eq:Btheta}, respectively.
We choose $\mu=1$ and $\epsilon=1, 10$ and $100$ in the axion insulator.
From Eq.~\eqref{eq:epsilontheta}, the corresonding values of $\Theta_0$ are $4.3\times 10^2, 1.4\times 10^3$ and $4.3\times 10^2$, respectively.
In the limit of $E_0/E_0^{\rm crit}\to \infty$, $|\theta|$ goes to $\pi/2$.
$E$ and $B$ in this limit are given by Eqs.~\eqref{eq:E_model1} and \eqref{eq:B_model1}, respectively.
Since the induced $\theta$ is much smaller than $\Theta_0$[region (I) of Fig. 3], the screening of $E$ at $E_0>E_0^{\rm crit}$ is too small
to be identified in the scale shown in (b).
}
}
\label{fig:dbt1}
\end{center}
\end{figure}

\revise{In Fig.\ref{fig:dbt1}, we show the induced $\theta$, $E$ and $B$ in this model, as functions of $E_0$.}
{We numerically find the position of the minima of Eq.~\eqref{eq:Hb_model1}, which is shown in Fig.~\ref{fig:dbt1}(a).
The corresponding $E$ and $B$ are obtained from Eqs.~\eqref{eq:Etheta} and \eqref{eq:Btheta}
and shown in Figs.~\ref{fig:dbt1}(c) and \ref{fig:dbt1}(d), respectively.
In particular, in the limit of $E_0/E_0^{\rm crit}\to \infty$, 
$\theta$ approaches to $\pi/2$ or $-\pi/2$ because of the divergence of the potential term at $\theta=\pm \pi/2$.
In this limit, $B$ and $E$ inside the axion insulator are linear in $E_0$: 
}
\revise{Because the induced $|\theta|\le\Theta$ in insulators, from Eqs.(\ref{eq:Btheta}) and (\ref{eq:Etheta}),
we can estimate $B$ and $E$ inside the material as }{}
\begin{align}
\lim_{E_0 \to \infty} B&=\tilde{B}\left(\pm\frac{\pi}{2}\right)\sim\pm3.6\times 10^{-3}\frac{\mu}{\epsilon}\epsilon_0E_0,
\label{eq:B_model1}\\
\lim_{E_0 \to \infty} E&=\tilde{E}\left(\pm\frac{\pi}{2}\right)\sim\left(1-1.3\times 10^{-5} \frac{\mu}{\epsilon}\right)\frac{\epsilon_0E_0}{\epsilon}.
\label{eq:E_model1}
\end{align}
\revise{which clearly show}{Equation~\eqref{eq:E_model1} clearly shows} 
that the electric field $E$ is only partially screened \revise{}{($1.3\times 10^{-5}\mu/\epsilon$)}.
While the massive axion electrodynamics analyzed by Ooguri and Oshikawa shows 
\revise{the complete screening of electric field beyond $E^{\rm crit}_0$ (see Appendix \ref{sec:ooguri-oshikawa}),  
no such a complete screening is seen in this model because large fluctuations of $\theta$ are strongly suppressed
in insulators with fixed $\phi_4$.}
{
screening of the excess electric field above $E^{\rm crit}_0$ (see Appendix \ref{sec:ooguri-oshikawa}), 
no such significant screening is seen in  this model because large induction of $\theta$ is strongly suppressed
in insulators with fixed $\phi_4$.
}

\section{Response to an applied magnetic field}
\label{sec:appliedB}
So far, an electric field {is} applied to the interface.
Now, we comment briefly what happnes when a magnetic field $B_0$, instead of $E_0$, is applied.

The boudary condition at the interface is given by
Eqs.~(\ref{eq:boundary}) and (\ref{eq:boundaryhap}).
Then, minimizing the Hamiltonian density (\ref{eq:hamiltonian}) with respect to $E$ under the boundary condition~(\ref{eq:boundaryhap}),
we obtain the solution $E=0$. 
With the optimized $B$ and $E$, 
the Hamiltonian density 
corresponding to Eq.
(\ref{eq:Hb_theta0}) can be written as
\begin{align}
\tilde{\mathcal H}_{\rm b}&=\frac{B_0^2}{8\pi\mu}
+\tilde{\mathcal V}_{\rm a}(\theta).
\end{align}
Hence, the induced axion field $\theta$ is determined by solving 
\begin{eqnarray}
\partial {\tilde{\cal V}}_{\rm a}/\partial \delta\theta=0,
\end{eqnarray}
and 
the effective square mass of axion  is obtained as
\begin{align}
M_{\rm eff}^2
=\left.\frac{\partial^2{\tilde{\mathcal V}_{\rm a}}(\theta)}{\partial\theta^2}
\right|_{\theta=0}{>0}.
\end{align}
Therefore, no axion instability happens and no electric field is induced inside the axion insulator. 

\section{Summary}
\label{sec:summary}
\revise{In this paper, we examine the axion instability in a microscopic point of view.
We introduce the antiferromagnetic field, instead of the axion field,
and analyze the instability caused by the antiferromagnetic field.
In contrast to the naive expectation, it is found that fluctuations of the antiferromagnetic field are insufficient to induce a visible axion instability.
We reveal that $\theta$ is bounded above as a function of the antiferrmagnetic field, and thus the instability is strongly suppressed.
In order to host the axion instability, fluctuations other than the antiferromagnetic field is necessary.
Analysis of quantum anomaly implies that the necessary fluctuation is related to topological quantum phase transition.
Only when fluctuations of the antimagnetic order and the topological quantum phase transition coexists, the axion instability occurs. }
{
In this paper, we examine axion instability from a microscopic point of view.
We introduce an antiferromagnetic field, instead of an axion field,
and analyze axion instability caused by the antiferromagnetic field.
From a general argument, 
it is pointed out that a non-magnetic order describing a topological transition is relevant to the axion dynamics, 
as well as the antiferromagnetic order.
Since an axion field is related to both magnetic and non-magnetic orders,
fluctuations of the antiferromagnetic field are insufficient to induce a visible axion instability.
}

\revise{}{
Starting from a microscopic Hamiltonian for a topological insulator with an additional term that breaks both time-reversal and inversion symmetries,
we  describe an axion field in terms of an antiferromagnetic field and an energy gap,
which correspond to the magnetic and non-magnetic orders, respectively.
Then we derive an effectiv Lagrangian for the axion field, $\theta$, and the electromagnetic fields, $E$ and $B$,
with a phenomenologically introduced potential for the axion field,
which is a $2\pi$-periodic function and includes higher order terms of $\theta$.
This potential keeps the system with $\theta=0$ at a local minimum of the energy density even under electromagnetic fields.
When an applied electric field exceeds a critical value, however, 
$\theta=0$ becomes an unstable point and the system relaxes to a new local minimum.
}

\revise{}{
To see the effect of the non-magnetic fluctuations,
we calculate induced $\theta$, $E$, and $B$ in response to an applied electric field for two model potentials with and without non-magnetic fluctuations.
In the case when both magnetic and non-magnetic fields fluctuate, a large amplitude of an axion field is induced above the critical field.
As a result, an applied electric field is almost completely screened.
Contrarily to this, in the case when only the magnetic order fluctuates, 
the amplitude of the induced axion field is bounded above by $\pi/2$, 
which cannot induce a significant screening of an electric field.
In both cases the induced magnetic field is small since it becomes significant only at around $\theta=\pi/\alpha\sqrt{\epsilon/\mu}$,
where $\epsilon, \mu$, and $\alpha$ are the dielectric constant and the magnetic permeability of the insulator and the fine-structure constant, 
respectively.
We also note that no axion instability occurs when a magnetic field, instead of an electric field, is applied.
Our result suggests that 
a system that is close to the topological phase transition point as well as the qunatum critical point of the antiferromagnetic order
is appropreate for invstigating axion electromagnetism.
}
}

\acknowledgements
The authors are grateful to K. Taguchi for valuable discussions.
This work is supported in part by a Grant-in Aid for Scientific Research from MEXT of Japan, ``Topological Materials Science,'' 
Grant No. 
JP15H05851, JP15H05853, and JP15H05855. 
M.S. is supported by Grant-Aid for scientific Research B (Grant No. JP17H02922) from JSPS.
Y.T. is supported by the Core Research for
Evolutional Science and Technology (CREST) of the Japan
Science and Technology Corporation (JST) [JPMJCR14F1].
\revise{}{Y.K. is supported by JST-CREST (Grant No. JPMJCR16F2), and JSPS KAKENHI Grant No. JP15K17726.}

\vspace*{3ex}
\appendix
\section{Ooguri-Oshikawa's theory}
\label{sec:ooguri-oshikawa}

\begin{figure}[h]
\begin{center}
\includegraphics[width=8cm]{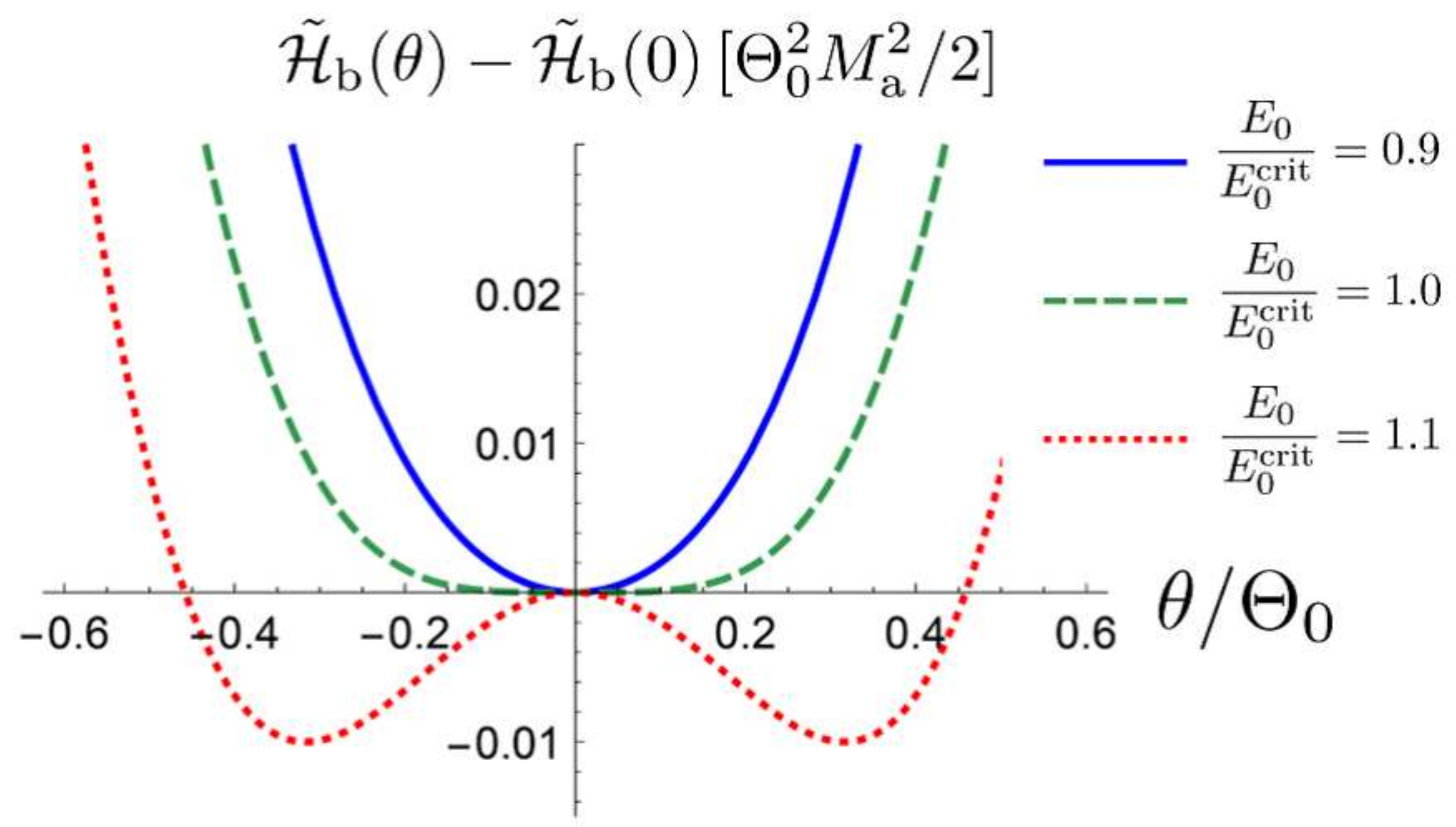}
\caption{
\revise{
$\tilde{H}_b(\theta)$ is plotted  as a function of $\theta$ based on Ooguri-Oshikawa's theory \cite{Ooguri12}.}
{
$\tilde{\mathcal H}_{\rm b}(\theta)$ based on Ooguri-Oshikawa's theory \cite{Ooguri12} is plotted  as a function of $\theta$.}
}
\label{fig:h_oo}
\end{center}
\end{figure}

\begin{figure}[h]
\begin{center}
\includegraphics[width=7cm]{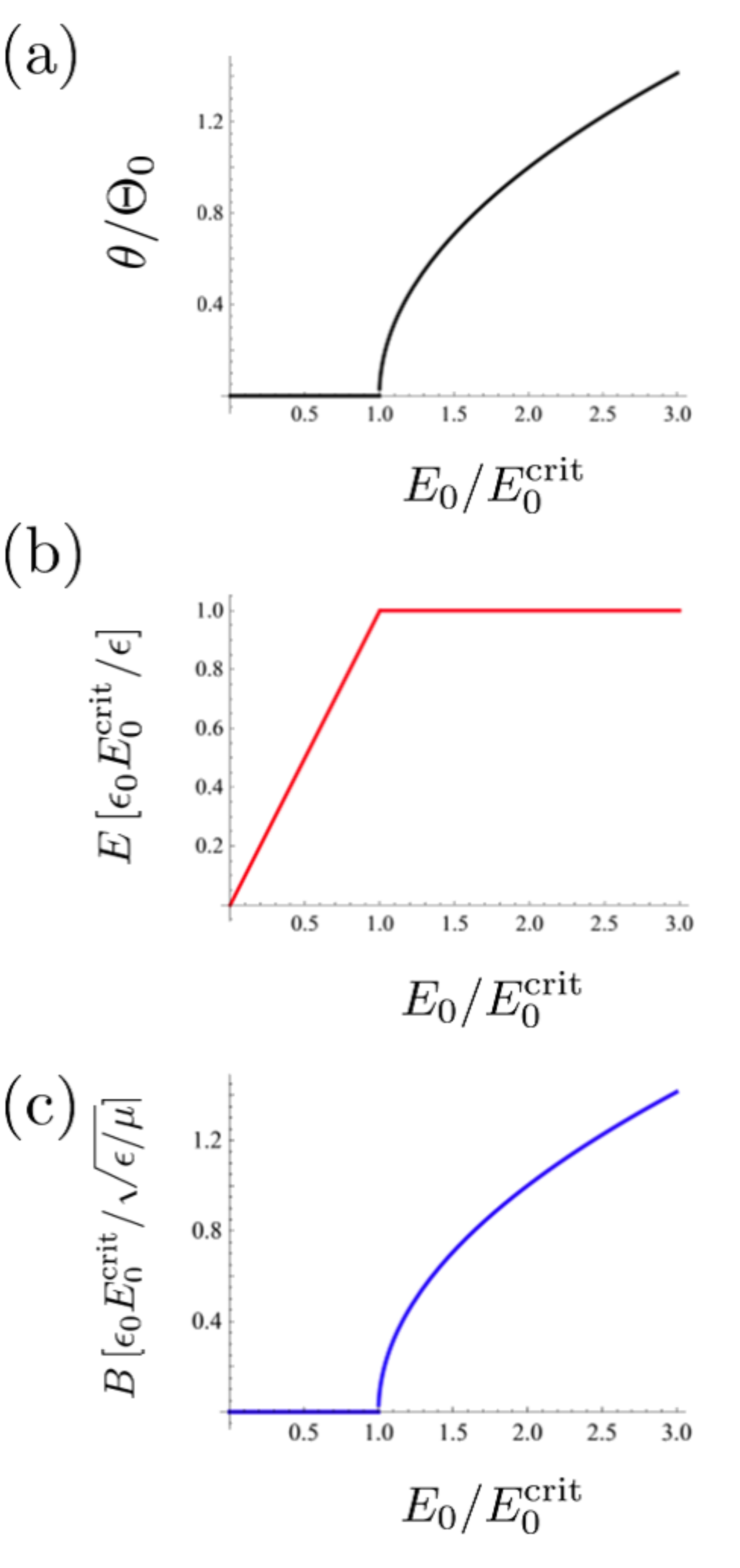}
\caption{
\revise{Applied electric field dependence of 
$\theta$, $E$ and $B$ based on Ooguri-Oshikawa's theory \cite{Ooguri12}. 
(a)$\theta/\pi$ is plotted as a function of $E_{0}$. 
(b)$\theta/\Theta_{0}$ is plotted as a function of $E_{0}$. 
(c)$E$ is plotted as a function of $E_{0}$. 
(d)$B$ is plotted as a function of $E_{0}$.}
{Behaviors of (a) the axion field $\theta$, (b) the electric field $E$, and (c) the magnetic flux density $B$
inside an axion insulator in response to an applied electric field $E_0$ based on Oogri-Oshikawa's theory\cite{Ooguri12},
where the values at $E_0>E_0^{\rm crit}$ are given by Eqs.~\eqref{eq:theta_oo}, \eqref{eq:E_oo}, and \eqref{eq:B_oo}, respectively.
The $\epsilon$ and $\mu$ dependences are all included in the scaling factor: $\Theta_0$ for $\theta$, $\epsilon_0E_0^{\rm crit}/\epsilon$ for $E$,
and $\epsilon_0E^{\rm crit}_0/\sqrt{\epsilon/\mu}$ for $B$.
Above the critical field, the electric field takes a constant value $\epsilon_0E_0^{\rm crit}/\epsilon$,
and a significant magnetic flux density is induced.
}\\
}
\label{fig:dbtoo}
\end{center}
\end{figure}

For comparison, we revisit the Ooguri-Oshikawa model.
They consider the following model potential, 
\begin{eqnarray}
\tilde{\mathcal V}_{\rm a}(\theta)=\frac{M_{\rm a}^2}{2}\theta^2, 
\label{eq:potential_oo}
\end{eqnarray}
which has  the minimum value $0$ at $\theta=0$. 
\revise{and satisfies Eqs.(\ref{eq:mass}) and (\ref{eq:massm5}).
The critical electric field $E_0^{\rm crit}$ is given by Eqs.(\ref{eq:Ecrit}) and (\ref{eq:mass2}).}{}
In the presence of \revise{the}{an} external electric field $E_0$ normal to the interface between the axion insulator and the normal insulator,
$\tilde{\mathcal H}_{\rm b}(\theta)$ is given by
\begin{align} 
&\tilde{\cal H}_{\rm b}(\theta)=\tilde{\cal H}_{\rm em}(0)\nonumber\\
&
+
\frac{M_{\rm a}^2\Theta_0^2}{2}
\left[-
\frac{(\theta/\Theta_0)^2}{1+(\theta/\Theta_0)^2}
\left(
 \frac{E_0}{E_0^{\rm crit}
}\right)^2
+
\left(\theta/\Theta_0\right)^2
\right]
.
\label{eq:Hb_oo}
\end{align}
\revise{See Fig.\ref{fig:h_oo}.}{The right-hand side of Eq.~\eqref{eq:Hb_oo} is shown in Fig.~\ref{fig:h_oo}.}
Then, $\partial \tilde{\mathcal H}_{\rm b}(\theta)/\partial \theta=0$ \revise{gives }{reduces to}
\begin{eqnarray}
\frac{1}{\theta/\Theta_0}
\left[
\frac{1}{(\theta/\Theta_0)^2}
-\left(\left|\frac{E_0}{E^{\rm crit}_0}
\right|-1\right)
\right]
=
0
.  
\label{eq:conditionofhb_oo}
\end{eqnarray}
\revise{From these equations, the induced $\theta$, $E$ and $B$ are obtained as}
{Above the critical electric field, $\tilde{\mathcal H}_{\rm b}(\theta)$ has two minima at}
\begin{align}
\theta
=\pm\Theta_0\sqrt{\left|\frac{E_0}{E^{\rm crit}_0}
\right|-1}.
\label{eq:theta_oo}
\end{align}
\revise{}{Substituting Eq.~\eqref{eq:theta_oo} in Eqs.~\eqref{eq:Btheta} and \eqref{eq:Etheta}, 
the induced $B$ and $E$ are obtained as}
\begin{eqnarray}
&&\revise{\theta
=\pm\Theta\sqrt{\left|\frac{E_0}{E^{\rm crit}_0}
\right|-1},}{}
\nonumber\\
&&\revise{\tilde{B}(\theta)}{B}
=\pm\frac{\epsilon_0E_0^{\revise{}{\rm crit}}}{\sqrt{\epsilon/\mu}}\sqrt{\left|\frac{E_0}{E^{\rm crit}_0}
\right|-1}
,
\label{eq:B_oo}\\
&&\revise{\tilde{E}(\theta)}{E}
=\frac{\epsilon_0E_0^{\rm crit}}{\epsilon}
.
\label{eq:E_oo}
\end{eqnarray}
We illustrate the induced $\theta$, $E$, and $B$ in Fig. \ref{fig:dbtoo} (a)-(d).
This model corresponds to the region (II) of Fig.~\ref{fig:db}.

In compariosn with the other  models considered in the main text, 
the Ooguri-Oshikawa model induces a larger magnetic field. \revise{when applying an electric field}{}
However, the justification of their analysis is not obvious. 
Their model potential takes into account only the squared term of $\theta$, and neglects the higher order terms. 
However, when a larger magnetic field is induced, $\theta$ becomes $O(1)$ so the higher order terms can not be neglected. 
Indeed, if $\theta$ originates from only magneric fluctuations,
our analysis  n Sec.~\ref{sec:magnetic_fluctuation} indicates that the axion instability should be suppressed due to the higher order terms.

%

\end{document}